\documentclass[pra,twocolumn,floatfix,showpacs,preprintnumbers,groupeaddress,amsmath,amssymb]{revtex4-1}
\usepackage[dvips]{graphics}

\usepackage[utf8]{inputenc}
\usepackage{mathptmx}
\usepackage{psfrag}
\usepackage{graphicx}						
\graphicspath{{figures/}}
\usepackage{braket}
\usepackage{rotating}
\usepackage{xcolor}
\definecolor{hblue}{RGB}{0,114,189} 
\usepackage{color}
\definecolor{dred}{rgb}{.6,.0,0.}
\definecolor{dblue}{rgb}{.0,.0,0.6}

\usepackage{tikz}
\usetikzlibrary{shapes.geometric}% für ellipse

\colorlet{mfarbe}{magenta}

\begin{document}

\title{Relaxation dynamics and dissipative phase transition in quantum oscillators with period tripling}

\author{Jennifer Gosner}
\author{Björn Kubala}
\author{Joachim Ankerhold}
\affiliation{Institute for Complex Quantum Systems and IQST, University of Ulm, 89069 Ulm, Germany}
\date{\today}

\begin{abstract}
Period tripling in driven quantum oscillators reveals unique features absent for linear and parametric drive, but  generic for all higher-order resonances. Here, we focus at zero temperature on the relaxation dynamics towards a stationary state starting initially from a domain around a classical fixed point in phase space. Beyond a certain threshold for the driving strength, the long-time dynamics is governed by a single time constant that sets the rate for switching between different states with broken time translation symmetry. By analyzing the lowest eigenvalues of the corresponding time evolution generator for the dissipative dynamics, we find that near the threshold the gap between these eigenvalues nearly closes. The closing becomes complete for a vanishing quantum parameter.
%we find that near the threshold an approximate closing of the gap between these levels occurs which becomes exact for a vanishing quantum parameter. 
We demonstrate that this behavior, reminiscent of a quantum phase transition, is associated with a transition from a stationary state which is localized in phase space to a delocalized one. We further show, that switching between domains of classical fixed points happens via quantum activation, however, with rates that differ from those obtained by a standard semiclassical treatment. As period tripling has been explored with superconducting circuits mainly in the quasi-classical regime recently, our findings may trigger new activities towards the deep quantum realm.
\end{abstract}

%pacs{73.23.Hk, 73.40.Gk, 72.70.+m, 85.35.Gv}

%
% 73.63.Kv Electronic transport in mesoscopic or nanoscale materials and
%          structures: Quantum dots
% 73.23.Hk Electronic transport in mesoscopic systems:
%          Coulomb blockade; single-electron tunneling
% 73.40.Gk Electronic transport in interface structures: Tunneling
%
% 73.23.-b 	Electronic transport in mesoscopic systems
% 72.70.+m 	Noise processes and phenomena
%		within electronic transport in condensed matter not in surfaces, films,...
%
% 42.50.Lc 	Quantum fluctuations, quantum noise, and quantum jumps
%		within quantum optics 
%
% 85.35.Gv 	Single electron devices

\maketitle

\section{Introduction}

A system in contact with a thermal reservoir relaxes when initially prepared in a state out of thermal equilibrium. This fundamental process is ubiquitous in all fields of science and is particularly interesting if, after a transient period of time, the dynamics is completely governed by only a single characteristic time scale, the inverse of the relaxation rate. The archetypical situation is that of an ensemble of particles initially confined in a potential well which is separated from an adjacent well or a continuum by a sufficiently high energy barrier. Then, the local dynamics within the well can be assumed to be much faster than barrier escape processes due to either thermal activation or quantum tunneling \cite{RevModPhys.62.251,Weiss2012}.
% P. H\''anggi, P. Talkner, and M. Borkovec, Rev. Mod. Phys. {\bf 62}, 251 (1990); U. Weiss, Quantum Dissipative Systems, World Scientific (Singapore, 2012).
 The situation becomes even richer when a system in contact with a thermal bath is externally driven periodically to approach a stationary state: First, these states can be of very diverse nature, for example, following the periodicity of the external source or not (broken time-translational symmetry) \cite{PhysRevLett.117.090402,PhysRevLett.109.160401,PhysRevLett.114.251603,PhysRevB.93.174305,PhysRevLett.116.250401,PhysRevB.93.245146,PhysRevLett.118.030401,Sacha_2017};
 % Dominic V. Else, B. Bauer,  C. Nayak (2016), Phys. Rev. Lett. 117, 090402 (2016). and others from the PRA
 second, the relaxation dynamics towards these states occurs not by traversing energy barriers but rather dynamical barriers in phase space. Accurate control of the parameters of the external drive (amplitude, frequency) allows one not only to precisely access different stationary states but also to explore the relaxation dynamics towards these states in different domains of phase space \cite{Kitagawa2012,PhysRevLett.106.220402,10.1038/nature21413,nature21426}.
 %experimental papers 

A particularly fascinating class of systems are periodically driven nonlinear oscillators that have received substantial attention in the last decade. Despite their putative simplicity, they reveal a wealth of dynamical features due to the subtle interplay of driving, nonlinearity, and dissipation and allow for a wide range of experimental implementations from superconducting circuits to nanomechanical systems, and  cold atomic gases \cite{Dykmanbook2012,PootvanderZant2012}. 
Mostly the conventional cases of linear \cite{DykmanKrivoglaz1980,PeanoThorwart20062,SerbanWilhelm2007,GuoZhengLi2010, GuoZhengLiZhengLiYan2011,andreguopeanoschoen2012} and parametric driving \cite{DykmanMaloneySilverstein1998,PhysRevB.83.224506,DykmanMarthalerPeano2011,PhysRevB.87.184501,doi10106315116533,PhysRevApplied.8.024018} of a weakly anharmonic oscillator have been studied. These systems are conveniently described in a rotating frame by quasi-energy Hamiltonians , where quasienergy levels are associated with Floquet states. Typically, the symmetry of  the Hamiltonian in the laboratory frame with respect to time translations is reflected  in phase space by rotational symmetries. Stationary states may then appear in the rotating frame as either localized or delocalized accessible by tuning drive parameters. The transition between them is generally associated with the occurrence of slow modes, bifurcations, and the existence of multiple orbits.

At low temperatures, quantum fluctuations are the dominant source to induce switching between classical fixed points.
Accordingly, when the quantum oscillator is initially prepared close to one of the classical fixed points, relaxation to the stationary state occurs either via quantum tunneling \cite{PhysRevA.76.010102, PhysRevLett.109.090401}, in absence or for very weak dissipation, or by quantum activation, for stronger dissipation \cite{MarthalerDykmanswitchin2006, DykmanMarthalerPeano2011,PhysRevLett.110.047001, Peano_2014}. The latter phenomenon is a manifestation of the fact that in a rotating frame excitation and relaxation processes between local quasi-energy levels behave very different from the situation for Fock states in the laboratory frame. Likewise, a stationary state is in general not determined by a detailed balance condition \cite{MarthalerDykmanswitchin2006, Peano_2014} in contrast to a thermal equilibrium. Additionally, time translation symmetry breaking has been found for period-two vibrations where the state of full symmetry (the unbroken state) can merge with the broken states \cite{PhysRevE.92.022105}.  In any case, driven nonlinear quantum oscillators may serve as testbeds to explore features of phase-transition-like phenomena far from equilibrium. 

In contrast to the cases of linear and parametric driving, much less attention has been paid to the study of period tripling \cite{nature21426,ZhangGosnerDykman2017,zhangDykman2019,1742-6596-681-1-012018,loerchDykman2019}, where the oscillator is periodically driven with three times its fundamental frequency. In fact, this is not just a generalization of the conventional situation but rather displays unique dynamical features, absent for linear and parametric drive, but generic for all higher order resonances \cite{PhysRevLett.111.205303,1367-2630-18-2-023006,doi10106315116533}. Classically, stationary orbits oscillating with an integer multiple of the drive period do {\em not} emerge continuously out of period-1 orbits (oscillating with the frequency of the drive). Quantum mechanically, tunneling between broken-symmetry states depends on a rotational geometric phase in phase space \cite{ZhangGosnerDykman2017}, very different from the case of the parametric oscillator \cite{PhysRevA.76.010102, PhysRevLett.109.090401}. Even weak dissipation destroys coherent tunneling and quantum activation is expected to set in.

While period tripling in the pure quantum case has been studied in \cite{ZhangGosnerDykman2017} and the quantum activation in \cite{zhangDykman2019}, here, we complement these findings by exploring in detail the relaxation dynamics towards stationary states and the dissipation-induced transition between these states being either localized or delocalized in phase space. Indeed, experimentally higher order photon resonances have already been observed in a set-up including superconducting resonators \cite{SvenssonShumeikoDelsing2017,180209259,SandboWilson2019}. They may be of relevance for the creation of higher order cat-states for quantum simulations, new types of quantum limited detectors, or to explore the fundamental physics of phase transitions far from equilibrium. Here, we lay the basis to explore the deep quantum regime and provide detailed predictions for future experiments, e.g. with superconducting circuits including Josephson junctions.

To set the stage, in Sec.~\ref{sec:clorbits} we introduce the generic model and briefly discuss the classical behavior in both the laboratory frame and the rotating frame. For the generalization to the quantum regime, we study the relaxation dynamics starting from a localized state in phase space numerically in Sec.~\ref{sec:simulation} based on a Lindblad-Master equation and in Sec.~\ref{escaperatelambda} in Liouville space for the dissipative generator of the dynamics. Section \ref{sc:semiclassical} collects some results for the switching in the semiclassical regime which allows in Sec.~\ref{sec:comparison} for a comparison of the various extracted relaxation rates and in Sec.~\ref{sec:tunneling} a comparison between ground state  quantum tunneling and quantum activation. Finally, Sec.~\ref{sec:transition} addresses the phase transition between localized and delocalized phase. Main findings are summarized in the Conclusions.

\section{Preliminaries: Classical nonlinear oscillator} 
\label{sec:clorbits}
We start by briefly addressing the classical regime which besides introducing the setting and the basic notation provides a physical picture to better understand the quantum problem.

\subsection{Steady state orbits}

\begin{figure}[htp]
\centering 
\includegraphics[width=1\linewidth]{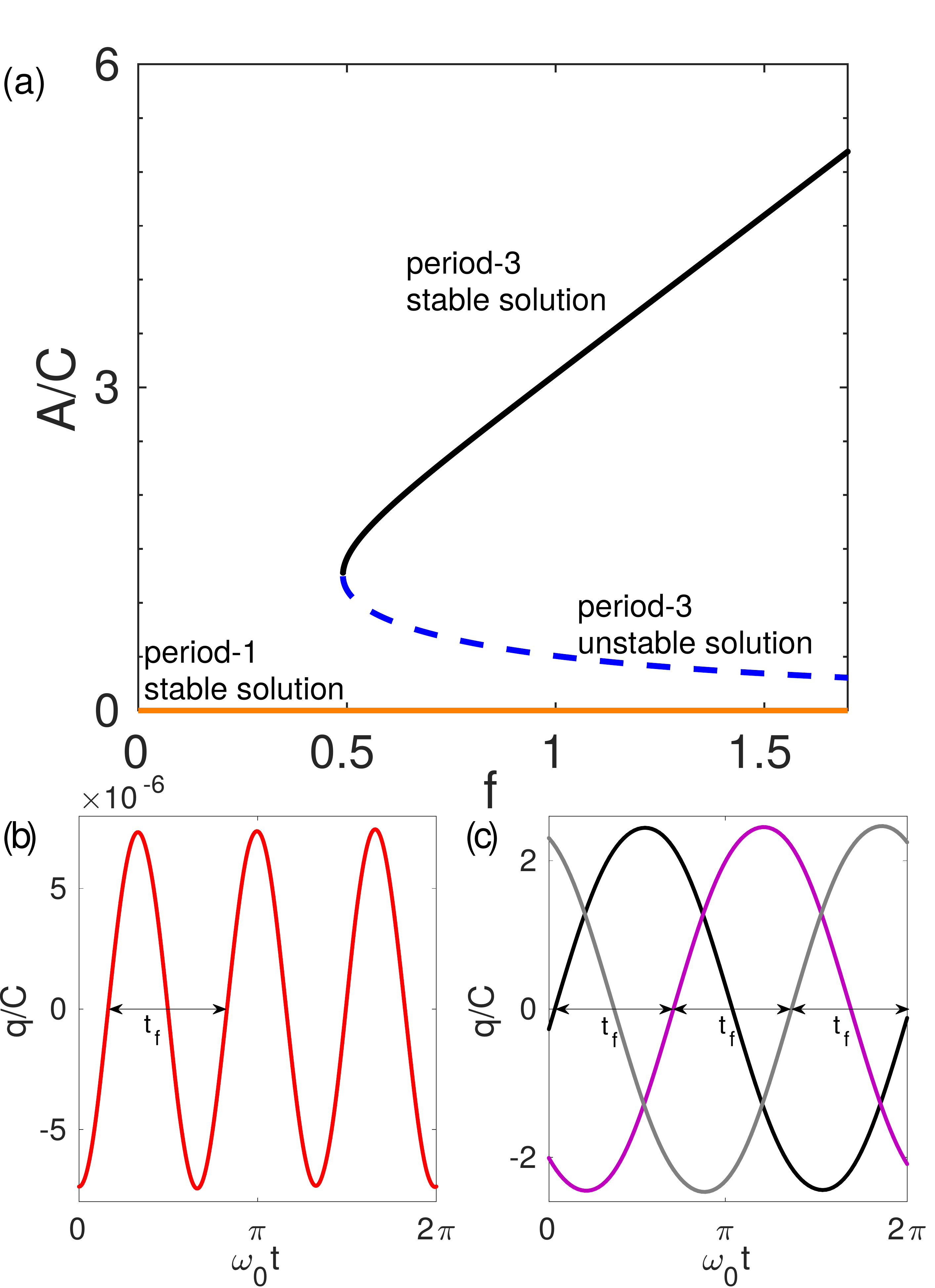}
 \caption{(a): Classical bifurcation diagram in rotating frame depicting steady state amplitudes [in units of $C=\sqrt{8 \omega_0 \delta \omega/3 \alpha}$] versus driving strength [scaled as in (\ref{scaleddriving})] for $\gamma/\omega_0=0.01$, $\omega_F/\omega_0=3.02$; the threshold beyond which multiple orbits coexist is $f_{c, \rm cl}\approx 0.49$. (b),(c): Steady state orbits in the laboratory frame solving \eqref{eqofmot} for $f=0.76$. The period-1 orbit (b)  is oscillating with the driving frequency $\omega_F$ (note the scale of the amplitude). Stable period-3 orbits (c) oscillating with $\omega_F/3 \approx \omega_0$  occur with relative phase-shifts $2\pi/3$, $4\pi/3$; 
 other parameters are $\alpha L^2/\omega_0^2=0.003$, $F_0 L/\omega_0^2=0.05$ with arbitrary length scale $L$.}
 \label{bifdiag}
\end{figure}

The system consists of a mechanical model, where a weakly anharmonic oscillator of the generic form
\begin{equation}
\ddot{q} + 2 \gamma \dot{q} + \omega_0^2 q + \alpha q^3 = F_0 q^{2} \cos(\omega_F t)
\label{eqofmot}
\end{equation}
is subject to weak damping with rate $\gamma$ and to an external driving with amplitude $F_0$ and frequency $\omega_F$. The nonlinearity is parametrized by $\alpha$ which is supposed to be weak, i.e., $(\alpha/\omega_0^2) A^2\ll 1$, where $A$ is a typical amplitude of steady state orbits (see Fig.~\ref{bifdiag}). For the driving frequency we will specifically focus on  $\omega_F \approx 3 \omega_0$ with a quadratic coupling to the oscillator degree of freedom, in contrast to the conventional situations of linear [$\omega_F \approx \omega_0$ with $F_0\cos(\omega_F t)$] and parametric [$\omega_F \approx 2 \omega_0$ with $F_0 q\cos(\omega_F t)$] driving, respectively. As we have demonstrated already in \cite{ZhangGosnerDykman2017}, this case is of particular interest as it shows features that are absent in these conventional situations but generic for all higher than second order resonances.

By assuming periodic steady state solutions of the form $q(t)=A \cos(\Omega t+\varphi)$ possible orbits can easily be obtained from (\ref{eqofmot}), for details see App.~\ref{appendix}. Two types of orbits are found, namely,  those oscillating with the  frequency $\Omega=\omega_F$ of the external drive, and those oscillating with $\Omega=\omega_F/3 \approx \omega_0$, the fundamental frequency of the bare oscillator; the former are termed period-1 orbits, the latter period-3 orbits. A stability analysis reveals that there is a stable and an unstable branch of period-3 orbits, see Fig.\ref{bifdiag}(a). Period-3 orbits only exist beyond a threshold for the driving strength $F_0$ and do not grow continuously out of period-1 solutions, in contrast to the situation for parametric amplification, where period-2 solutions smoothly emerge out of period-1 orbits.
While the amplitudes for stable period-3 orbits grow with increasing driving, those for unstable period-3 orbits shrink and asymptotically approach the stable branch of the period-1 orbits. Amplitudes for the latter always remain small due to the far-off-resonance driving, see Fig.\ref{bifdiag}(b). %
According to the time translational symmetry of (\ref{eqofmot}), period-3 solutions appear (up to an overall off-set) with three different phases $\varphi=0, 2\pi/3, 4\pi/3$, see Fig.\ref{bifdiag}(c). Note that if initially one prepares the oscillator in a period-1 state for weak driving and adiabatically increases the driving, in absence of external perturbations, the oscillator never approaches period-3 orbits. 

\subsection{Quasi-energy in the rotating frame} 
The non-dissipative part of (\ref{eqofmot}) results from the time-dependent Hamiltonian
\begin{equation}
H=\frac{p^2}{2}+\frac{\omega_0^2 q^2}{2}+ \frac{\alpha q^4}{4} - \frac{F_0 q^3}{3} \cos(\omega_F t)\, .
\end{equation}
If the driving is not too strong, so that the driving term, the anharmonic term, and $\omega_0 \delta\omega q^2$ with  de-tuning $|\delta\omega|=|\omega_F/3-\omega_0|\ll \omega_0$ are small compared to the bare harmonic part, it is convenient to map this Hamiltonian in the laboratory frame via a canonical transformation of the form
\begin{equation}
q  = C Q \cos\left(\frac{\omega_F t}{3}\right) + C P \sin\left(\frac{\omega_F t}{3}\right)
\label{unitarytrafo}
\end{equation}
to a rotating frame Hamiltonian $H_{RWA}$. Here, we introduced dimensionless quadratures $Q$, $P$ and  a scaling factor $C=\sqrt{\frac{8 \omega_0 \delta \omega}{3 \alpha}}$ (we assume $\delta\omega>0$ in the sequel). This way, one finds the dimensionless quasi-energy
\begin{equation}
\label{quasienergy}
g(Q,P)\equiv \frac{8 H_{RWA}}{3 \alpha C^4}=\frac{1}{4} (Q^2+P^2-1)^2-  f (Q^3-3 P^2 Q), 
\end{equation}
with a dimensionless driving strength
\begin{equation}
\label{scaleddriving}
f=\frac{F_0}{3\sqrt{24 \omega_0 \alpha \delta \omega}}\, . 
\end{equation}
	 \begin{figure}[htp]
\centering 	
\includegraphics[width=1\linewidth]{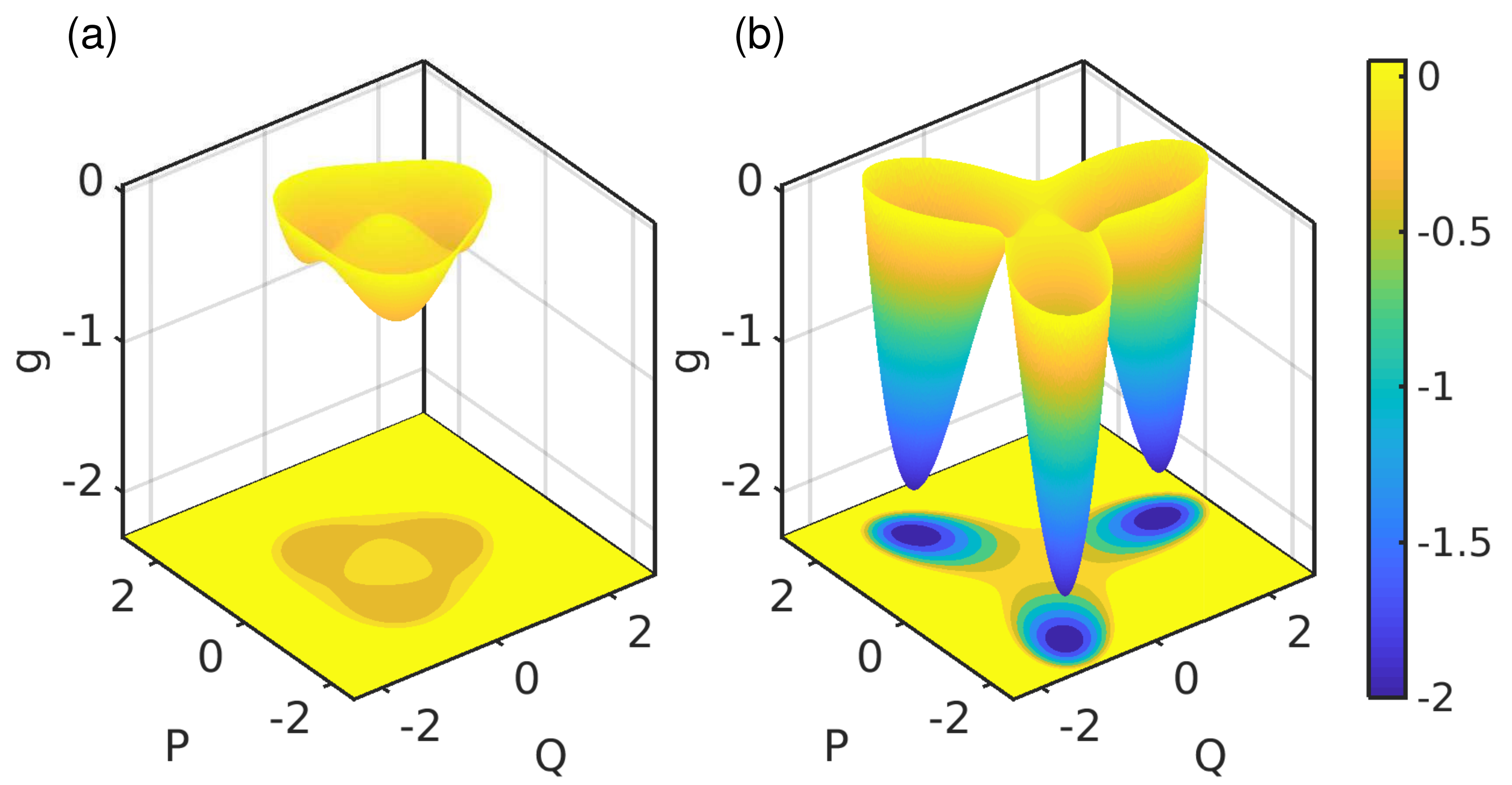}
\caption{Classical quasi-energy $g(Q,P)$ according to (\ref{quasienergy}) for different driving strengths $f=0.1$ (a) and $f=0.5$ (b). The quasi-energy has a $C_3$ symmetry in $QP-$ space, where the local maximum at $Q_0=0, P_0=0$ corresponds to the period-1 orbit in Fig.~\ref{bifdiag}(b) and the three wells located at $(Q_m,P_m)$, $m=1,2,3$ are related to the period-3 orbits in Fig.~\ref{bifdiag}(c). }
\label{classpotg}
\end{figure}
Note that here the definition of $f$ differs by a numerical factor from \cite{ZhangGosnerDykman2017, zhangDykman2019}.
This quasi-energy has a three-fold symmetry in $QP-$ space ($C_3$ symmetry), see Fig.~\ref{classpotg}, that reflects the time-translational symmetry of the three steady state period-3 orbits discussed in the previous section. Accordingly, a local maximum at $Q_0=0, P_0=0$ associated with the period-1 orbit with low amplitude is complemented by three wells located at $(Q_m,P_m)$, $m=1,2,3$ corresponding to the period-3 orbits with
\begin{equation}
\begin{split}
&P_1=0 \text{,  } Q_1=Q_{\rm min}\\
&P_{2,3}= \pm \sin(2\pi/3)\,  Q_{\rm min} \text{,  } Q_{2,3}=\cos(2\pi/3)\,  Q_{\rm min}\, ,
\end{split}
\end{equation}
where $Q_{\rm min}=\frac{3}{2}f + \sqrt{\frac{9}{4}f^2+1}$  (see \cite{ZhangGosnerDykman2017}). Rotations in the $QP$-plane  by $2\pi/3$ correspond  to time translations in the laboratory frame $t\to t+t_F, t_F=2\pi/\omega_F$, so that rotations by $2\pi$ map onto $t\to t+3t_F$, see \cite{ZhangGosnerDykman2017} for further details. The three wells are separated by saddle points with quasi-energy $g_{\rm saddle}<0$. The equations of motion in the rotating frame in presence of dissipation easily follow from Hamilton's equations augmented by friction, i.e.,
\begin{equation}
\begin{split}
&\dot{Q}= \partial_P g - \kappa\,  Q\\
&\dot{P} = - \partial_Q g - \kappa\,  P
\label{eq:classical}
\end{split}
\end{equation}
with the dimensionless friction constant $\kappa=\frac{\gamma}{\delta \omega}$. In App.~\ref{appendix} we discuss a protocol how to realize experimentally the situation in (\ref{eqofmot}) via a conventional Duffing oscillator with linear drive. We note in passing that time translational symmetry breaking in the classical realm has  recently been studied experimentally in a two-oscillator set-up in \cite{heugel2019}.
%heugel2019, Toni L. Heugel, Matthias Oscity, Alexander Eichler, Oded Zilberberg, R. Chitra, arXiv: 1903.02311 (2019)
%

\section{Dynamics towards steady state in the quantum regime}
\label{sec:simulation}

We will now turn to the main subject of this paper, namely, the quantum induced switching out of domains in $QP$-space around classical fixed points. The quantum equivalent to the canonical transformation \eqref{unitarytrafo} is provided by the unitary operator ${U}=\exp({-i \frac{\omega_F t}{3} {a}^{\dagger} {a}})$ with the standard ladder operators of the bare harmonic system obeying $[a, a^\dagger]=1$.   The quantized rotating frame Hamiltonian is then obtained as
\begin{equation}
\begin{split}
\hat{g}=& - \Lambda n + \frac{1}{4}\left(1-\Lambda\right)^2 + \Lambda^2 (n+n^2) - 4 f \left(\frac{\Lambda}{2}\right)^{\frac{3}{2}} (a^3+ a^{\dagger 3} )\, 
\end{split}
\label{quasi-operator}
\end{equation}
with the effective Planck constant 
\begin{equation}
\label{lambda}
\Lambda=\frac{ \hbar}{M \omega_0 C^2}\equiv \frac{3\hbar\alpha}{M 8\omega_0^2\delta\omega} \,,
\end{equation}
including the mass $M$ of the oscillator. The effective Planck constant serves as an externally tunable parameter, for example via the de-tuning $\delta\omega$. 
The operator $\hat{g}$ has an equivalent representation in terms of phase space operators $\hat{Q}, \hat{P}$ with canonical commutation relations  $[\hat{Q},\hat{P}]=i \Lambda$ according to ${a} = \frac{1}{\sqrt{2 \Lambda}} ({\hat{Q}} + {i \hat{P}})$ and ${a}^{\dagger} = \frac{1}{\sqrt{2 \Lambda}} (\hat{Q} - {i \hat{P}})$. This way, $\hat{g}(\hat{Q}, \hat{P})$ results from the classical quasi-energy $g(Q,P)$ in (\ref{quasienergy}) by replacing classical variables by operators and $P^2 Q \to \hat{Q}\hat{P}\hat{Q}$. 

For the dissipative dynamics of the quantum oscillator we invoke a weak coupling approximation which leads to a standard
Lindblad master equation
\begin{equation}
\begin{split}
\dot{\rho}(t) =& \frac{1}{i \Lambda} [g,\rho] +  \kappa (\bar{n}+1) \left(2a \rho a^\dagger -  a^\dagger a \rho -  \rho  a^\dagger a\right)\\
&+ \kappa \bar{n} \left(-2a^\dagger \rho a + a a^\dagger \rho +  \rho a a^\dagger\right)\, ,
\end{split}
\label{Lindblad}
\end{equation}
where time is scaled with $\delta \omega$ and $\bar{n}=1/[\exp(\hbar \omega_0\beta)-1]$ at inverse temperature $\beta=1/k_{\rm B}T$ denotes the Bose distribution.

With the above ingredients and starting from a specific initial state, one can now explore the quantum dynamics analogous to the discussion of the classical steady state orbits above. Here, we are particularly interested in the quantum fluctuations of one of the wells around $(Q_m, P_m), m>0$ in the limit of vanishing temperature $\bar{n}=0$. For that purpose, we consider the dynamics of a locally relaxed state namely the stationary solution of the local Fokker-Planck equation in the harmonic well around $Q_1$, $P_1$ \cite{Dykmanbook2012,PhysRevE.75.011101}. This way we obtain in the Wigner representation
\begin{equation}
\rho_W(Q, P,t=0) \propto \exp\left[- \frac{2}{\Lambda} \left(\frac{g_{QQ} \delta Q^2 + g_{PP} \delta P^2} {g_{PP}+g_{QQ}} \right) \right]\, ,
\end{equation}
 where $\delta Q=Q-Q_1$, $\delta P=P-P_1$, and $g_{XY}$ is the second derivative with respect to $X, Y$ taken at $Q_1, P_1$. 
 
\begin{figure}[htp]
\centering  	 
\includegraphics[width=1\linewidth]{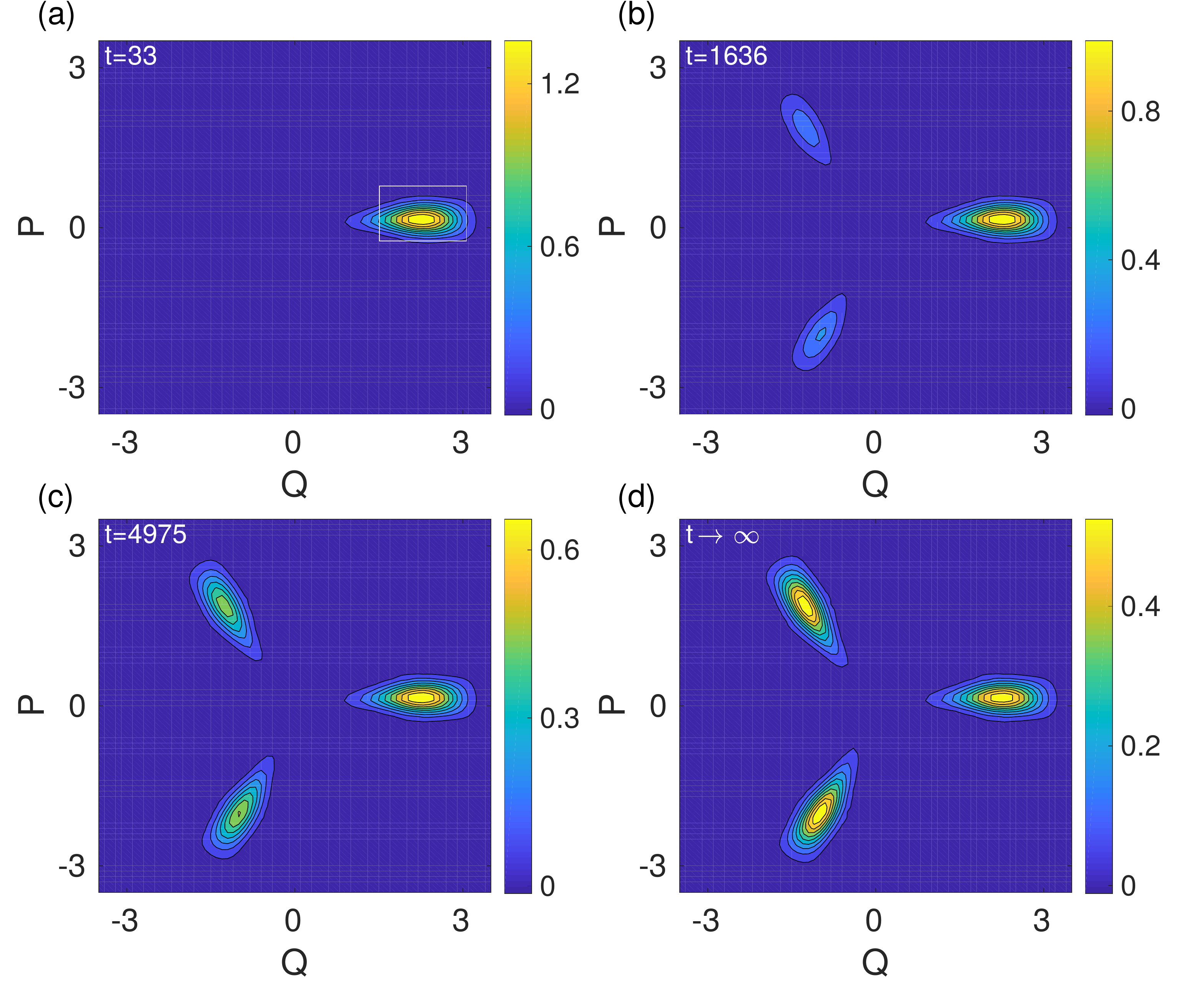}			
 \caption{Wigner distribution $\rho_W(Q,P,t)$ starting initially from a locally relaxed state in the well $Q_1, P_1$ of the quasi-energy surface at different time steps and for $\kappa=1.5$, $\Lambda=0.1688$, $f=0.9$: $t=33$ (a), $t=1636$ (b), $t=4975$ (c), and asymptotic state (d). The population dynamics of a closed area indicated by a white box in (a) is depicted in Fig.~\ref{decayswitching}, such that the population $p_1(0)\approx 1$ at time $t=0$.}
 \label{Wignerlindz}
\end{figure}

% We monitor the dynamics of this initial state according to (\ref{Lindblad}) with normalization
% \begin{equation}
%\int \frac{dQ dP}{2 \pi \Lambda}\,  \rho_W(Q,P,t) =1 \, . 
%\end{equation}
%Snapshots of corresponding Wigner distributions at different time steps are shown in Fig.~\ref{Wignerlindz}. 
Monitoring the dynamics of this initial state according to (\ref{Lindblad}) we show snapshots of corresponding Wigner distributions at different time steps in Fig.~\ref{Wignerlindz}. Apparently, in the long time limit the distribution approaches a steady state that is delocalized predominantly among the three wells with minimal contributions around the origin. The set of parameters is chosen such that classically the system is beyond the threshold (see Fig.~\ref{bifdiag}), where period-1 and period-3 orbits coexist. This in turn implies that quantum fluctuations induce the decay of classically stable fixed points towards delocalized distributions with $C_3$-symmetry. Note the slight distortions of the arms of the distribution due to friction.

\begin{figure}[htp]
\centering  	 
\includegraphics[width=1\linewidth]{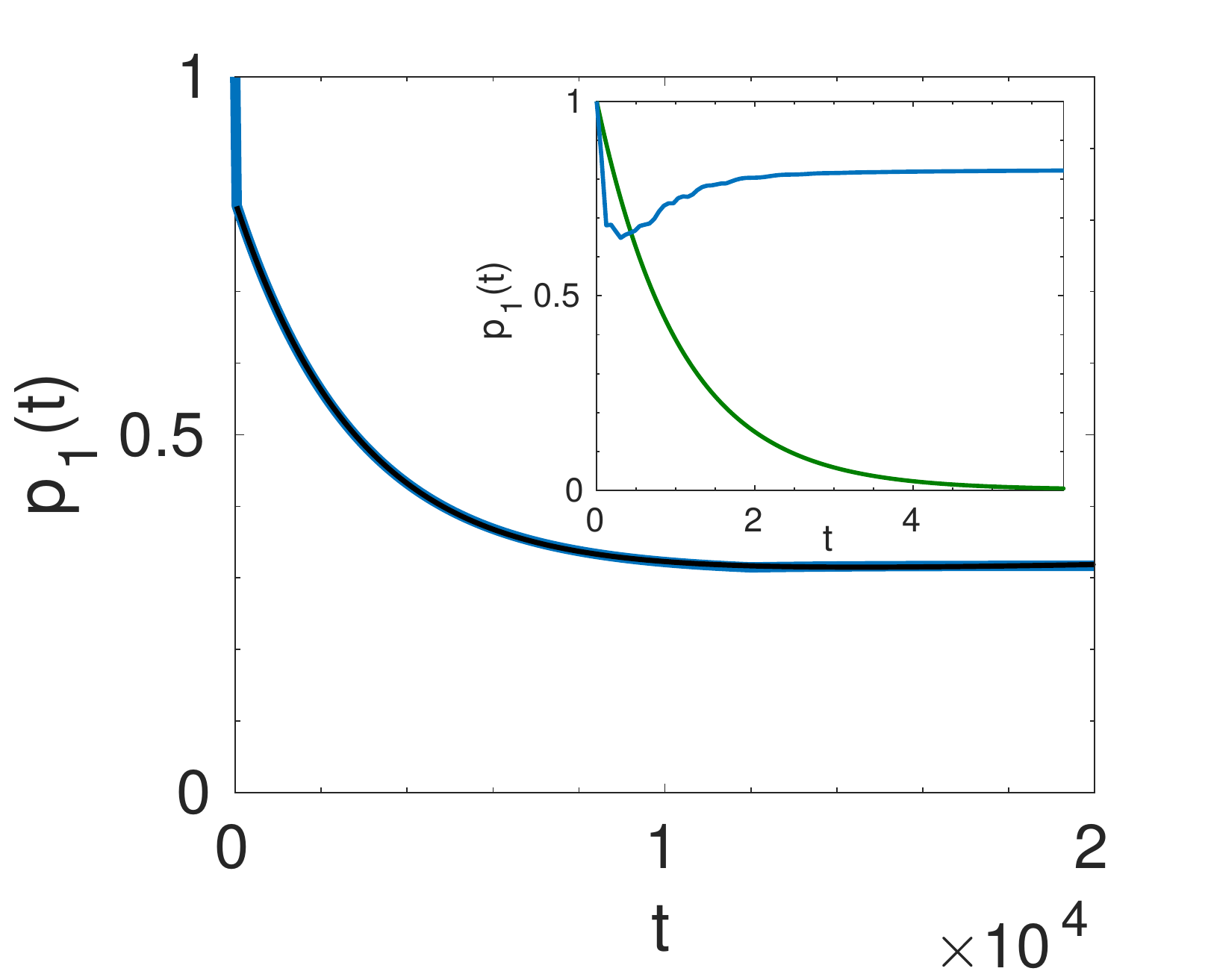}			
 \caption{Decay of the population in the well around $Q_1, P_1$ (in the white box in Fig.~\ref{Wignerlindz}(a)) versus dimensionless times for $\kappa=1.5$, $\Lambda=0.1688$, $f=0.9$. On long time scales it is governed by a single time constant $1/\Gamma_L$. The black line is a fit function, where $p_1(\infty)=0.32$. The inset shows the transient behavior on very short time scales (blue); also shown is the decay of a state initially located around the origin at $Q=P=0$ (green). The initial distribution at $Q=P=0$ decays much faster than the initial population around the well $Q_1, P_1$.}
 \label{decayswitching}
\end{figure}

After a transient period of time, this relaxation is very accurately characterized by a single time scale $\Gamma_L$ when following the population dynamics
\begin{equation}
p_1(t) = \int_{B_1} dQ dP \, \rho_W(Q,P,t)\, 
\end{equation}
according to $\dot{p}_1(t)=- \Gamma_L [p_1(t) - p_1(\infty)]$, see Fig.~\ref{decayswitching}. Here, $B_1$ indicates a closed area in $QP$-space around $Q_1, P_1$ (white box in Fig.~\ref{Wignerlindz}(a)) such that initially $p_1(0)\approx 1$. Asymptotically, one has $p_1(\infty)<1/3$ due to the portion of the distribution that is still located around $Q=P=0$. 
%Apparently, this is relaxation is in contrast to coherent oscillations between the wells that are expected to occur in complete absence of dissipation and on different time scales (cf.\ also below in Sec.~\ref{sec:tunneling}).
Apparently, the observed dynamics is purely relaxational in contrast to coherent oscillations between the wells that are expected to occur in complete absence of dissipation and on different time scales (cf.\ also below in Sec.~\ref{sec:tunneling}). As we discussed in \cite{ZhangGosnerDykman2017}, due to the energy level structure of (\ref{quasi-operator}) sufficiently beyond the threshold, only exponentially weak friction is needed to induce decoherence (associated with transitions between tunnel-splitted energy levels) and mixing of corresponding eigenstates. For friction that  substantially exceeds the tunnel splitting (as considered here) the coherent dynamics turns into uni-directional decay. Thus, on the time scale associated with $\Gamma_L$ there occur transitions between the localized wells  (interwell transitions) that, as we will discuss in detail in Secs.~\ref{sc:semiclassical} and \ref{sec:tunneling}, are attributed to the phenomenon of quantum activation. A condition for the long time dynamics to be governed by this single time scale is a time scale separation between interwell and intrawell processes,  an issue that we will explore more carefully in the next section. 

Furthermore, we show in Fig.~\ref{decayswitching} that starting initially from a ground state distribution localized around $Q=P=0$ (classical period-1 orbit) provides a much faster decay towards the delocalized asymptotic steady state. % as depicted in Fig.~\ref{decayswitching}. 
The time scale associated with this decay turns out to play an important role near those domains in parameter space, where the nature of the steady state changes from localized to delocalized, i.e. the quantum analogue of the classical bifurcation as will be discussed in Sec.~\ref{sec:transition}.

\section{Time scale separation between intrawell and interwell processes} 
\label{escaperatelambda}

In the previous section, the time scale which governs the long time dynamics was obtained from the time evolution of the Lindblad master equation $\dot{\rho}(t)=\mathcal{L} \rho(t)$ with the super-operator $\mathcal{L}$ as given in (\ref{Lindblad}). Here, we extend this analysis to access also shorter time scales which capture not only inter-well but also intra-well dynamics. The most convenient way to do so, is to switch to a Liouville representation, where the density is considered as a vector in Liouville space on which  $\mathcal{L}$ operates.  The goal is then to find eigenfunctions and eigenvalues of  $\mathcal{L}$ according to
\begin{equation}
\mathcal{L} \ket{\ket{v_k}} = \lambda_k \ket{\ket{v_k}} ,
\end{equation}
where the density matrix at time $t=0$ is represented as
\begin{equation}
\ket{\ket{\rho(0)}}=\sum_{k\geq 0} a_k \ket{\ket{v_k}}\, .
\end{equation}
Of course, due to the existence of a steady state $\mathcal{L} \rho_{ss}=0 $ there is always a vanishing eigenvalue $\lambda_0=0$; all other eigenvalues have negative real parts ${\rm Re}\{\lambda_k\}<0$ and are either real or appear in complex conjugate pairs. The Lindblad time evolution in Liouville space
\begin{equation}
\ket{\ket{\rho(t)}}= e^{\mathcal{L} t} \ket{\ket{\rho(0)}}=a_0 \ket{\ket{v_0}}+\sum_{k\geq 1} a_k e^{\lambda_k t} \ket{\ket{v_k}}  ,
\end{equation}
is completely determined by the initial state projections $a_k$ and the eigenvalues $\lambda_k$,  and asymptotically approaches $\rho_{ss}= \lim\limits_{t \rightarrow \infty}{\rho(t)}\sim \ket{\ket{v_0}}$ .

In the sequel, we focus at zero temperature $\bar{n}=0$ on the lowest eigenvalues. The Liouville operator $\mathcal{L}$ is converted accordingly to a $N^2\times N^2$ matrix which is diagonalized using the Lanczos/Arnoldi algorithm. 
This provides eigenvectors  $\ket{\ket{\nu_l}}, l\geq 0$ to each eigenvalue $\lambda_l$ which are linear independent, but not orthogonal.

Upon representing eigenvectors $\ket{\ket{\nu_l}}$ again in $N\times N$ matrices, i.e.\ $\ket{\ket{\nu_l}}\to \rho^{(l)}$, one can classify the eigen-densities into three classes according to the following structure in the Fock state basis
\begin{eqnarray}
\ket{\ket{\nu_{3k}}}\to \rho^{(3k)} &=& \sum_{n \ge 0}^N p_{nn}^{(k)} \ket{n} \bra{n} \nonumber\\
&&+ \sum_{n,m\ge 0}^N \left(\alpha_{nm}^{(k)} \ket{n} \bra{3m+3+n}+h.c.\right) \nonumber\\
\ket{\ket{\nu_{3k+1}}}\to \rho^{(3k+1)} &=&  \sum_{n,m\ge 0}^N \Big(\gamma_{nm}^{(k)} \ket{n} \bra{3m+1+n}\nonumber\\&& + \delta_{nm}^{(k)}  \ket{3m+2+n} \bra{n}\Big)\nonumber\\
\ket{\ket{\nu_{3k+2}}}\to  \rho^{(3k+2)} &=&{\rho^{(3k+1)}}^\dagger \ \ , \ \ k\geq 0
\end{eqnarray}
with real-valued coefficients $p_n^{(k)}$ and complex-valued $\alpha_{nm}^{(k)}, \gamma_{nm}^{(k)}, \delta_{nm}^{(k)}$. Eigen-density matrices have matrix elements $\rho_{nm}$ where $n-m$ is identical $(\text{mod } 3)$ within the same class. This structure is explained by the fact that both the Hamiltonian \eqref{quasi-operator} part as well as the dissipative part of the Lindblad master equation \eqref{Lindblad} only couple such matrix elements. Diagonal elements of $\rho$ (populations) only appear in the class $\rho^{(3k)}$ associated with real eigenvalues $\lambda_{3k}$, while $\lambda_{3k+1}=\lambda_{3k+2}^*$. As we will see in Sec.~\ref{sec:transition}, for sufficiently weak driving the dominating contribution to the steady state density $\rho^{(0)}$ is the ground state density of the oscillator around $Q_0, P_0$, while for stronger driving beyond a threshold strongly delocalized  components prevail.
	 \begin{figure}[htp]
\centering  
\includegraphics[width=0.5\textwidth]{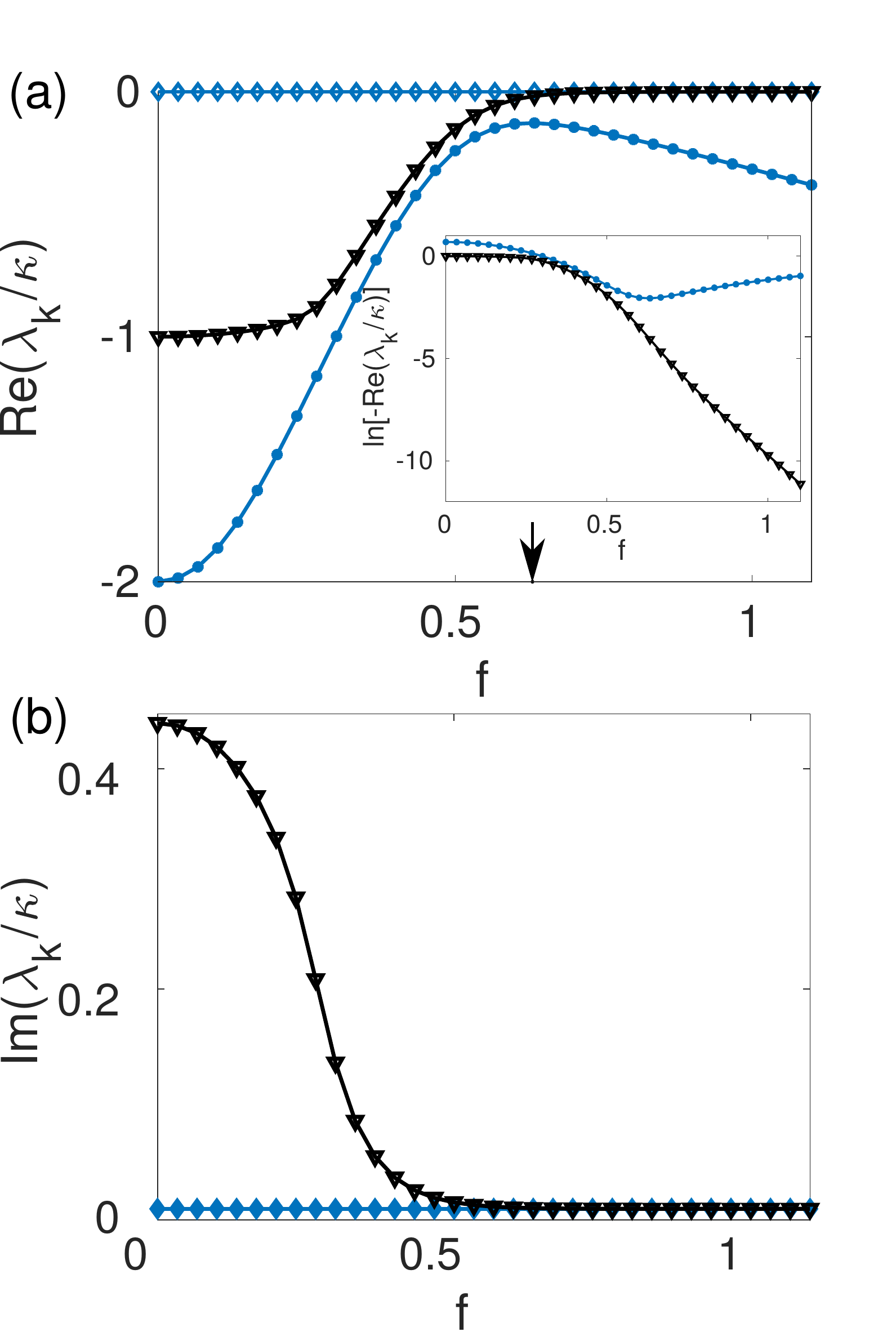}
 \caption{Real (a) and imaginary (b) parts of the lowest lying eigenvalues of the Lindbladian in (\ref{Lindblad}) versus the driving for $\Lambda=0.1688$ and $\kappa=1.5$. Shown 
 are eigenvalues corresponding to the class $\rho^{(3k)}$ (blue), namely, 
  $\lambda_0$ (\textcolor{hblue}{$\diamond$}) and $\lambda_3$ (\textcolor{hblue}{$\bullet$}), and eigenvalues corresponding to the classes $\rho^{(3k+1)}, \rho^{(3k+2)}$ (black), namely, ${\rm Re}\{\lambda_1\}={\rm Re}\{\lambda_2\}$ ($\nabla$). For very weak driving
  ${\rm Re}\{\lambda_{1, 2}\}$ describe relaxation of a harmonic oscillator near $Q=P=0$, while beyond a threshold $f_c$ (here at the maximum of $\lambda_3$) the eigenvalues become exponentially small (see inset of (a) with a log-scale) and capture interwell transition processes. The imaginary part (b) of $\lambda_{1,2}$ also decreases towards $\lambda_0$.}
\label{realimageigliou}
\end{figure}

Here, we proceed with the analysis of the subset of eigenvalues with least negative real parts.
Figure \ref{realimageigliou}(a) depicts together with $\lambda_0=0$ the three eigenvalues ${\rm Re}\{\lambda_1\}={\rm Re}\{\lambda_2\}, \lambda_3$ versus the driving strength. For very weak driving these eigenvalues reflect  the expected results for a harmonic oscillator located at $Q_0=P_0=0$, i.e.\ $\lambda_{1,2}=- \kappa$, $\lambda_3 = - 2 \kappa$, according to the low energy properties of $\hat{g}$ in (\ref{quasi-operator}) for $f=0$.
 With increasing driving, however, near a critical driving $f_c$, $ {\rm Re}\{\lambda_{1, 2}\}$ turn into an exponential decrease towards $\lambda_0=0$ for $f>f_c$ [see Fig.~\ref{realimageigliou}(a), inset]; a similar behavior is found for their imaginary parts [Fig.~\ref{realimageigliou}(b)]. The appearance of a new (exponentially small) time scale is associated with the appearance of a new dynamical process, namely, the switching between wells separated by quasi-energy barriers (dynamical barriers in phase space). For the present case, it is also accompanied by the tendency of closing of the gap between $\lambda_0=0$ and the next lowest lying eigenvalue $\lambda_3$, a signature typically associated with the emergence of slow modes well-known for phase transition phenomena \cite{RevModPhys.69.315}. 
 % S. L. Sondhi, S.M. Girvin, J. P. Carini, and D. Shahar, Rev. Mod. Phys. {\bf 69}, 315 (1997).
 It becomes an exact closing only for $\Lambda\to 0$ which may thus be interpreted as the thermodynamic limit for this system, see also Sec.~\ref{sec:transition} and Fig.~\ref{changeinlambda}. For any finite $\Lambda$ the transition is smeared out in a range around $f_c$.
 
Thus, we identify a critical driving $f_c\approx 0.63$  as that driving strength, where $|\lambda_3|$ approaches a minimum; as we will see in Sec.~\ref{sec:transition}, around this point the steady state changes from being localized around the origin of the $QP$-plane to being predominantly localized around $Q_m, P_m, m=1, 2, 3$. Of course, there is no sharp transition for finite $\Lambda$.  All other eigenvalues $\lambda_{k>3}$ remain sufficiently separated from $\lambda_3$ near $f_c$. Beyond this threshold, $|\lambda_3|$ increases again to become well-separated from ${\rm Re}\{\lambda_{1, 2}\}$ and dives down into the pattern of the other eigenvalues $\lambda_{k>3}$. For increasing $f>f_c$ this pattern approximately captures that of a harmonic oscillator localized in one of the wells $Q_m P_m, m=1, 2, 3$ (not shown). We thus conclude that for $f>f_c$, one has a clear time scale separation between intrawell and interwell processes so that the long time relaxation behavior is governed by a single time scale, i.e.\ $|{\rm Re}\{\lambda_{1, 2}\}|=\Gamma_L$. Near the critical driving $f_c$, another dynamical role is assigned to the eigenvalues $\lambda_{1,2}$: While for very weak driving they are related to the relaxation dynamics of the local harmonic oscillator at the origin of the $QP$-plane, for $f$ sufficiently beyond $f_c$ their real parts correspond to interwell transition processes. How this switching between classical period-3 states happens is in more detail elucidated in Secs.~\ref{sc:semiclassical} and \ref{sec:tunneling}. A general theory connecting metastability and separations in the eigenvalue spectrum of Markovian open quantum systems has been put forward in \cite{PhysRevE.94.052132,PhysRevLett.116.240404}.

We found a slightly modified pattern of the lowest lying eigenvalues for parameters $\kappa<\Lambda$, which, however, does not change the conclusions about the transition. Specifically for those parameters the eigenvalues exhibit a more complex dependence on the driving below $f_c$ and the critical behavior is observed for driving values slightly below the minimum of $|\lambda_3|$. For driving strengths near and above $f_c$ one finds a similar behavior as for $\kappa>\Lambda$ though.
%The above pattern of the lowest lying eigenvalues slightly changes for parameters, where $\kappa<\Lambda$, but qualitatively stays the same: Below $f_c$, eigenvalues exhibit a more complex dependence on the driving and the critical driving moves towards values slightly below the minimum of $|\lambda_3|$. For driving strengths near and above $f_c$ one finds a similar behavior as for $\kappa>\Lambda$ though.

\section{Semiclassical switching rate}
\label{sc:semiclassical}
In order to obtain a deeper insight into the interwell relaxation process, we employ a semiclassical treatment applicable when formally Planck's constant $\Lambda \ll 1$. Practically, this implies a large number of states in the potential wells of $g(Q, P)$ such that  $|g(Q_m, P_m)|/\Lambda \sim f^4/\Lambda\gg 1$ and also that $\kappa$ by far exceeds the tunnel splitting and switching rate to ensure a time scale separation (see below).
Technically, we follow the methodology explained in \cite{MarthalerDykmanswitchin2006}: One represents the master equation  \eqref{Lindblad} around one of the well basins in corresponding intrawell eigenstates (Wannier basis) $|\mu\rangle$ of $\hat{g}$ that are linear combinations of the global eigenstates, see \cite{ZhangGosnerDykman2017}. Since we are interested in the population dynamics, only the diagonal part $\rho_\mu\equiv \langle \mu|\rho|\mu\rangle$ is considered. We assume a quasi-continuum of energy levels in the wells so that quantum mechanical matrix elements can be obtained from classical orbits. Within this semiclassical approximation, the steady state distribution is obtained based on an exponential ansatz. For a system in thermal equilibrium this would lead to a Gibbs distribution, while here we find an exponent that depends nonlinearily on quasi-energies. Following these lines, we start from
\begin{equation}
\label{eq:lindblad2}
\dot{\rho}_{\mu}  = -2\kappa \sum_\nu \left(W_{\mu \nu}\rho_\mu - W_{\nu \mu}\rho_\nu\right)\,   
\end{equation}
with transition rates
\begin{equation}
W_{\mu+\nu, \mu}=(1+\bar{n}) |\langle \mu|a|\mu+\nu\rangle|^2 +\bar{n} |\langle\mu+\nu| a|\mu\rangle|^2\, .
\end{equation}
For vanishing temperature $\bar{n}=0$ only transitions from $\mu+\nu \rightarrow \mu$ take place. Semiclassically, the relevant matrix elements follow from 
\begin{equation}
\begin{split}
a_\nu(g_\mu) & \equiv \sqrt{2 \Lambda} \bra{\mu} a \ket{\mu+\nu}\\
&=\frac{1}{T(g)} \int_{P.O.} dt\,   {\rm e}^{- i \omega_\mu  \nu t}  [Q(g_\mu,t)+i P(g_\mu,t)]
\end{split}
\end{equation}
 where we set $(g_{\mu+\nu}-g_\mu)/\Lambda\approx \nu \omega(g_\mu)$ with $g_\mu=\langle \mu|\hat{g}|\mu\rangle$. The integration is taken along periodic orbits (P.O.) in the well area with position $Q(g,t)$ and momentum  $P(g, t)$ at fixed energy $g$ with period $T(g)=2\pi/\omega(g)$;  these are period-3 orbits in the language of Sec.~\ref{sec:clorbits}. Note that since the right hand side of (\ref{eq:lindblad2}) is of order $\kappa$, the orbits are obtained from (\ref{eq:classical}) in absence of friction. 
 
\begin{figure}[htp]
\centering  	 
\includegraphics[width=1\linewidth]{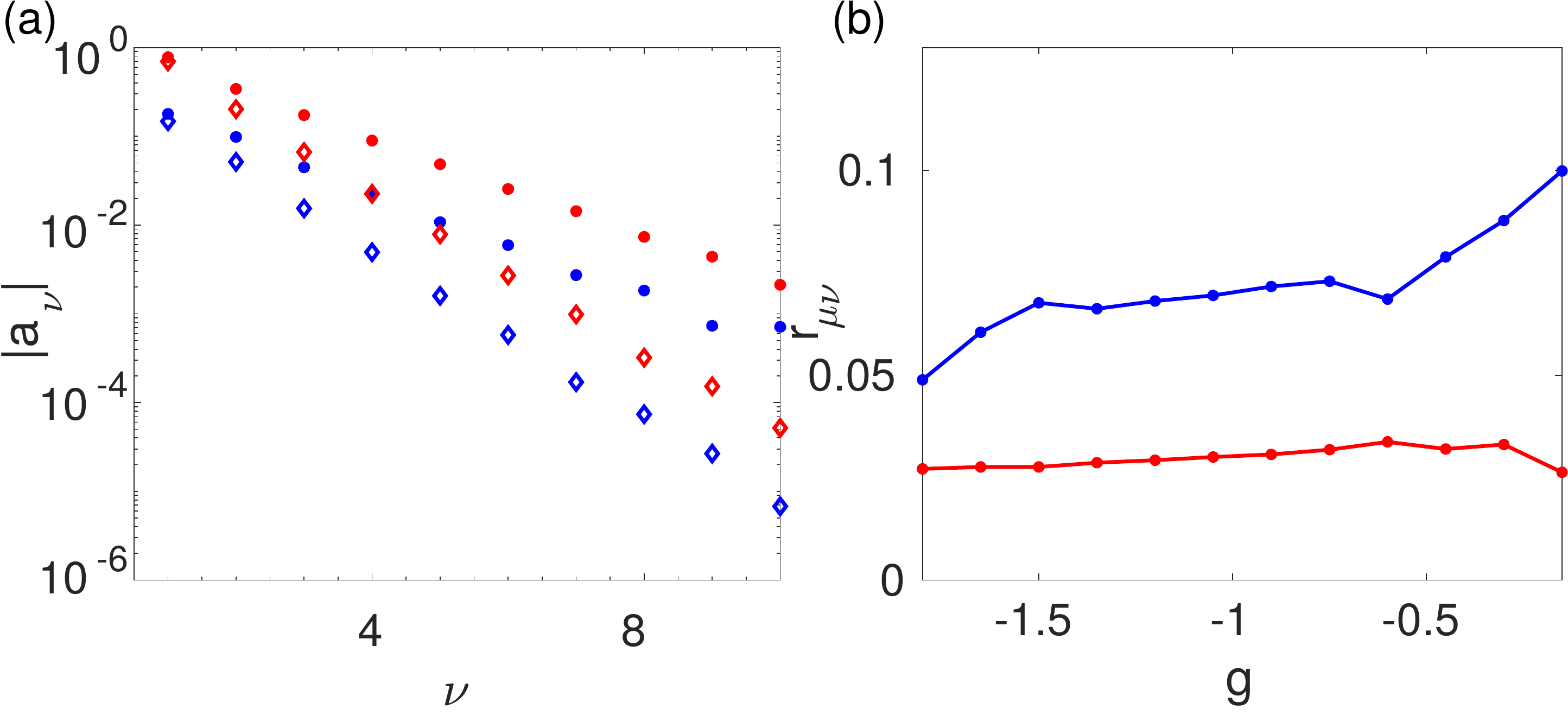} 	
 \caption{(a): Transition matrix elements $|a_\nu(g)|$ at two different quasi-energies $g=-1.2$ (diamonds), $g=-0.2$ (dots) for $f=0.66$ [$g_\text{min}=-3.55$, $g_\text{saddle}=0.22$]. Shown are $|a_\nu|$ (blue) and $|a_{-\nu}|$ (red), where $|a_\nu|<|a_{-\nu}| (\nu>0)$. The matrix elements decrease exponentially with growing $\nu$ and increase as energy grows towards $g_\text{saddle}$. (b): Detailed balance ratio $r_{\mu\nu}$ according to (\ref{detailedratio}) for $r_{42}$ (blue) and $r_{31}$ (red) at $f=0.5$. The ratio $r_{\mu\nu}$ differs from $1$ in both cases so that the detailed balance condition is not fulfilled.}
\label{asymmetricmatrixel}
\end{figure}

The above matrix elements carry interesting information about intrawell processes such as relaxation and quasi-energy diffusion in the rotating frame as we will discuss now. 
Figure \ref{asymmetricmatrixel}(a) depicts $a_\nu(g)$ versus $\nu$  at different values of $g$, where in accordance with the semiclassical limit we assume a quasi-continuum of energies $\{g_\mu\}\to g_{\rm min}\leq g\leq g_{\rm saddle}$. The general tendency is that the matrix elements decrease exponentially with growing $\nu$ and increase as the energy grows towards the saddle point.  Further, $|a_\nu|<|a_{-\nu}| (\nu>0)$ with a slight asymmetry in the decay as $\nu$ increases (for details and analytical results cf.~\cite{zhangDykman2019}). This unusual behavior at zero temperature is a manifestation of the fact that relaxation processes in the laboratory frame appear in the rotating frame as both excitation and relaxation processes which gives rise to the phenomenon of quantum activation \cite{MarthalerDykmanswitchin2006}.
Further, the asymmetry in the decay with growing $\nu$  has direct consequences on the detailed balance condition
\begin{equation}
\frac{W_{\mu\nu}}{W_{\nu\mu}}=\frac{W_{\mu\mu'} W_{\mu'\nu}}{W_{\nu\mu'} W_{\mu'\mu}}\, .
\end{equation}
As seen in Fig. \ref{asymmetricmatrixel}(b), it is not fulfilled , even for $\bar{n}=0$, since the ratio
\begin{equation}
\label{detailedratio}
r_{\mu\nu}=\frac{|a_\nu|^2 |a_{-\nu+\mu} a_{-\mu}|^2}{|a_{-\nu}|^2 | a_{\nu-\mu} a_{\mu}|^2}
\end{equation}
clearly differs from 1. This in turn is a signature of the fact that in the rotating frame dissipation induced relaxation and excitation processes behave quite differently to the situation in the laboratory frame. Notably, while for the Duffing oscillator with parametric drive, detailed balance is at least guaranteed at zero temperature, we here find detailed balance to be violated for {\em all} temperatures. 

Now, in steady state according to an eikonal ansatz the density is of the form $\rho_\mu\sim \exp(-R_\mu/\Lambda)$ with $R_\mu=R(g_\mu)$ so that $\rho_{\mu+\nu}\approx \rho_\mu \exp[-\nu \omega_\mu R'(g_\mu)]$,  $R'=dR/dg$. The balance equation (\ref{eq:lindblad2}) can then be cast in the form
\begin{equation}
\label{stat-master}
\sum_{\nu=-\mu}^{\mu_\text{saddle}}= W_{\mu+\nu,\mu} \left(1-\xi^\nu_{(\mu)} \right)=0, 
\end{equation}
where the parameter $\xi_{(\mu)} = e^{-R'_\mu\omega_\mu}$ is independent of $\Lambda$ in leading order. 
The above equation can easily be solved numerically for the parameter $\xi$ from which $R'(g)$ is obtained. This way, 
  by assuming a simple relation between steady state density and interwell relaxation rate (similar to the relation between the Boltzmann distribution and the thermal escape rate for barrier escape problems in absence of driving), one finds
\begin{equation}
\Gamma_{\rm scl} = D_0 \,  {\rm e}^{-\frac{R(g_{\rm saddle})}{\Lambda}}
\label{escsem}
\end{equation}
with an unknown prefactor $D_0$ and the action $R(g)=\int_{g_{\rm min}}^{g} dx R'(x) $, known for quantum activated processes \cite{MarthalerDykmanswitchin2006}. The calculation of the prefactor is much more challenging. According the known procedure for static barriers, one would seek in the low viscosity limit for a flux solution which describes deviations from the stationary distribution for energies close to the barrier top. The matching between this solution and the stationary one then provides the prefactor. It is not clear yet how such a procedure can be extended to phase space barriers and the deep quantum regime. The classical Kramers result \cite{RevModPhys.62.251} suggests $D_0\approx \kappa$ for sufficiently weak friction, while for somewhat larger friction 
$D_0\approx 1/T(g_{\rm min})$ is the expected transition state theory (TST) result. This issue will be discussed in the next section. 

One important remark is in order here: The basic assumption underlying the above treatment, particularly (\ref{stat-master}), is that state populations remain stationary due to transitions between neighboring states. As a recent study revealed \cite{zhangDykman2019}, however, this is in a strict sense no longer true for the present case, but only applies approximately for values of $f$ sufficiently away from $f_c$. In fact, as shown by Zhang and Dykman, stationarity is maintained by non-local transitions so that, at least in principle, one must start from the full balance equation (\ref{eq:lindblad2}) to determine the stationary distribution. Nevertheless, in case of a sufficient time scale separation between intrawell processes (fast) and interwell decay (much slower), the dominating contribution to the interwell rate (i.e.\ on a logarithmic scale) is still expected to be fairly accurately captured by (\ref{escsem}).

\begin{figure}[htp]
\centering 
\includegraphics[width=1\linewidth]{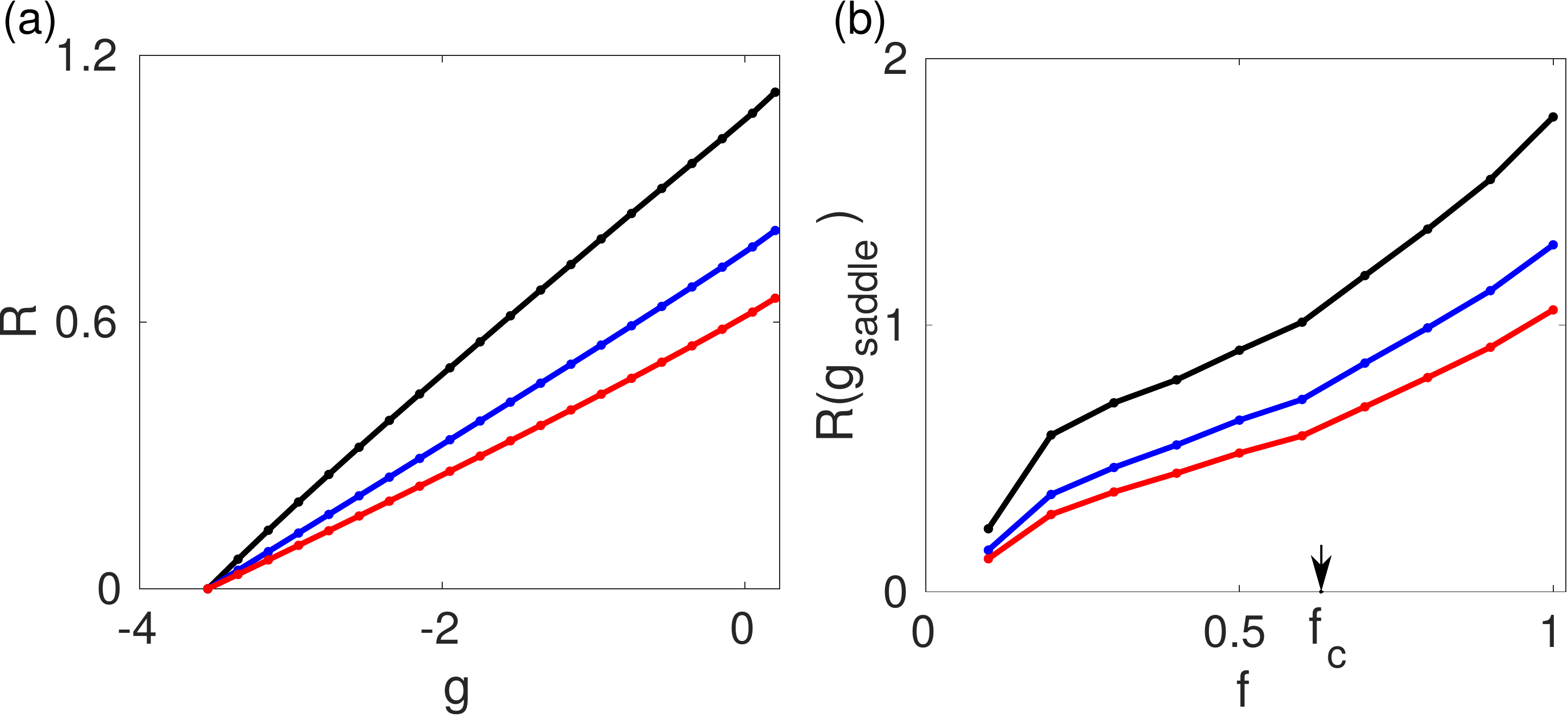}
\caption{(a): Action $R(g)$ at $f=0.66$ [$g_{\rm min}=-3.55, g_{\rm saddle}=0.22$] and for different temperatures: $\bar{n}=0$ (black), $\bar{n}=0.1$ (blue), $\bar{n}=0.2$ (red). The action increases linearly with $g$. (b): Saddle point action $R(g_{\rm saddle})$ determining the rate for quantum activation versus the driving strength at $\bar{n}=0$ (black), $\bar{n}=0.1$ (blue), $\bar{n}=0.2$ (red). $R(g_{\rm saddle})$ increases almost linearly with increasing driving beyond $f_c$. }
\label{RandRprime}
\end{figure}

Figure \ref{RandRprime}(a) illustrates the dependence of  the action $R(g)$ on $g$ for various temperatures $\bar{n}$.
The action grows basically linearly with $g$ with slight deviations for elevated temperatures. The action for interwell transitions $R(g_{\rm saddle})$ is shown in Fig. \ref{RandRprime}(b) versus the driving strength and reveals a monotonously growing, nonlinear behavior which turns into an almost linear one beyond $f_c$.

\section{Comparison of switching rates}
\label{sec:comparison}

	 \begin{figure}[htp]
\centering 
\includegraphics[width=1\linewidth]{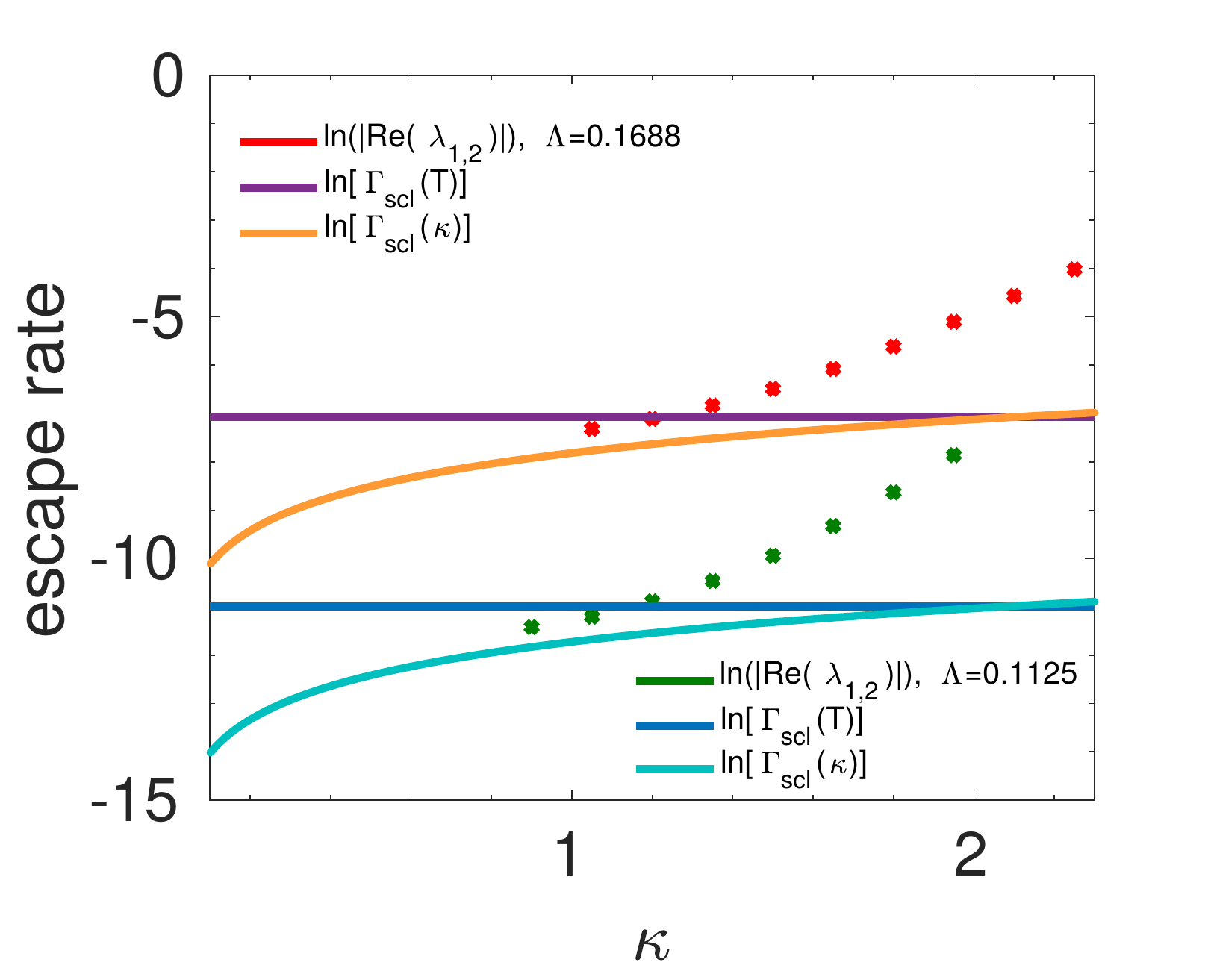}	
\caption{Semiclassical rate for interwell transitions via quantum activation $\Gamma_{\rm scl}$ according to (\ref{escsem}) with $D_0= 1/T(g_{\rm min})$ (purple and blue) and $D_0= \kappa$ (orange and turquoise) for $\Lambda=0.1688$ and $\Lambda=0.1125$ together with the numerically obtained switching rates $\Gamma_L=|{\rm Re}\{\lambda_{1, 2}\}|$  at $\Lambda=0.1688$ (red crosses) and  at $\Lambda=0.1125$ (green crosses) versus the friction constant $\kappa$  at $f=0.8$.}
\label{kappadepf}
\end{figure}

We now collect the results of the previous sections with respect to the relaxation time scale in the regime $f>f_c$. First, as to be expected, one finds that numerically to very high accuracy
\begin{equation}
\Gamma_L\approx | {\rm Re}\{\lambda_{1, 2}\}|, 
\end{equation}
for driving strengths sufficiently beyond the threshold, cf.~Fig.~\ref{realimageigliou}(a). The rate exponent depends roughly linearly on $f$ for $f>f_c$ [see inset of Fig.~\ref{realimageigliou}(a)], a behavior that is approximately also found for the exponent in $\Gamma_{\rm scl}$, see  Fig.~\ref{RandRprime}(b). However, while $\Gamma_L$ exhibits a dependence on the friction strength $\kappa$, this is by construction not the case for $\Gamma_{\rm scl}$ (see Fig.~\ref{kappadepf}) if $D_0 \propto 1/T(g_{\rm min})$ is chosen. Instead for $D_0 \propto \kappa$ we find a decreasing rate with decreasing $\kappa$ with substantial deviations, however, for growing $\kappa > 1$ (see Fig.~\ref{kappadepf}). Note that for the parameters chosen in Fig.~\ref{kappadepf} one has $V_b/\Lambda \omega_1 \approx 12$ ($\Lambda=0.1688$) or $\approx 18$ ($\Lambda=0.1125$) levels in the well ($V_b$ is the well depth and $\omega_1$ the frequency near the well bottom). The diagonalization of the full Lindbladian according to Sec.~\ref{escaperatelambda} is limited to values of the friction with $\kappa>0.9$. Nonetheless, the dominating exponential in $\Gamma_L$ is captured by the simple semiclassical expression $\Gamma_{\rm scl}$, while a more refined description is required for the prefactor. This must also include the impact of the period-1 oscillator located around $Q=P=0$ and the dynamics close to barrier energies.

\section{Quantum activation versus quantum tunneling} 
\label{sec:tunneling}
The question whether switching between the wells happens via quantum activation only or to which extent direct low-energy quantum tunneling between the wells (ground state tunneling between period-3 orbits) may play a role, has recently been addressed in \cite{zhangDykman2019}. Here, we recall the main results to keep the line of reasoning self-contained.
Accordingly, we consider also the dominant exponential factor for coherent tunneling from one well to the others (corresponding to the maximal  level splitting between adjacent energy levels of $\hat{g}$). In a semiclassical limit, the latter is given by $\propto e^{-2S_{tun}/\Lambda}$, where up to corrections of order $\Lambda$, the action at a given quasi-energy $g<g_{\rm saddle}$ reads
\begin{equation}
S_{\rm tun}(g)=\int_{Q_1}^{-Q_1/2} {\rm Im}\{P(Q)\} dQ\, 
\end{equation}
Note that since the integration contour includes not only areas, where $g<g(Q,P)$, the momentum carries apart from imaginary also real parts. Accordingly, the full action has also a real part that induces a phase dependence of the energy level splittings, for details see \cite{ZhangGosnerDykman2017}. 
 
\begin{figure}[htp]
\centering  	 
\includegraphics[width=1\linewidth]{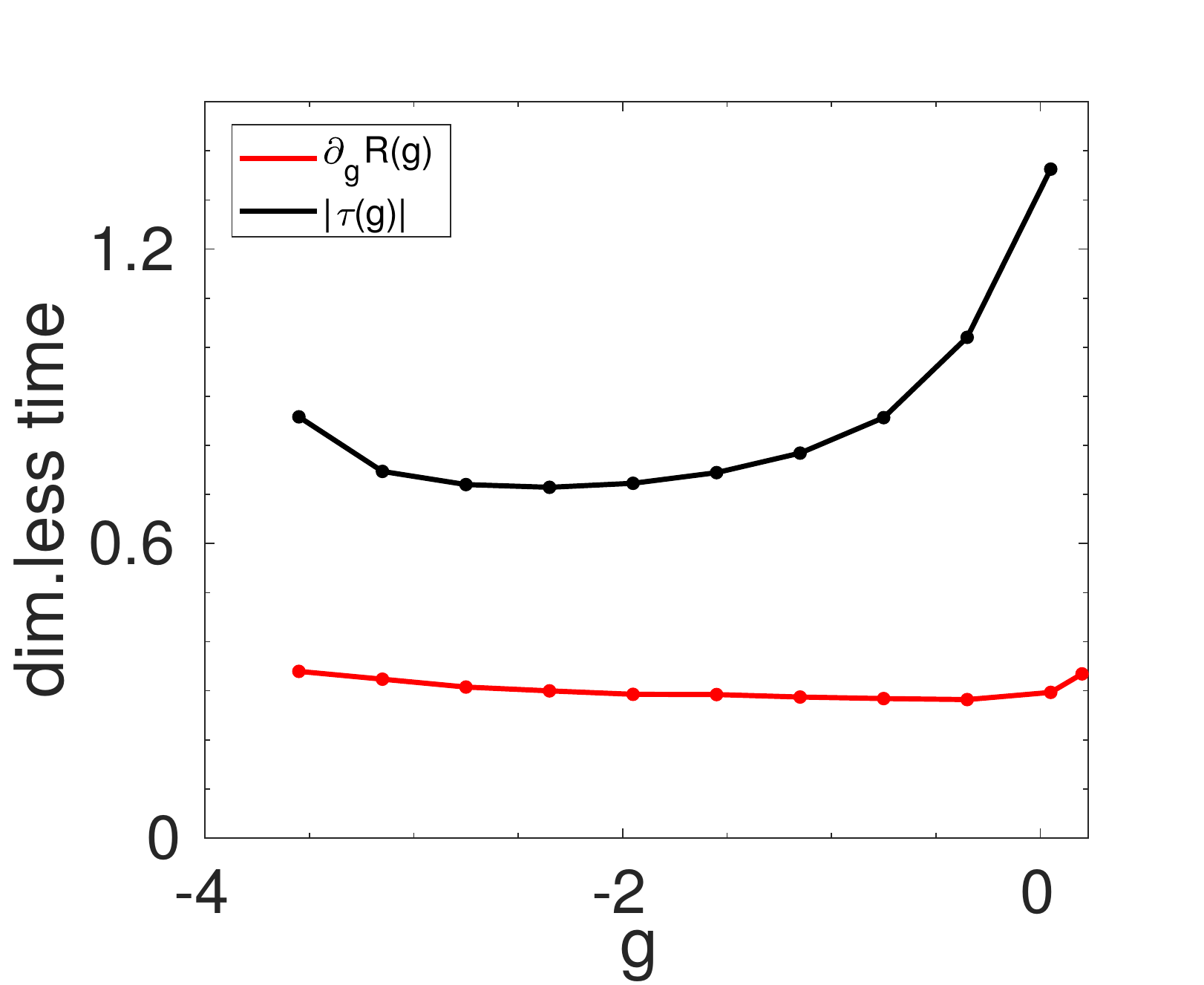}			
 \caption{Condition for quantum activation: The tunneling time $|\tau(g)|$ exceeds the slope of the action $R(g)$ for all energies $g$ at $f=0.66$ [$g_{\rm min}=-3.55, g_{\rm saddle}=0.22$] so that the condition is always fulfilled. Consequently, the relaxation dynamics towards a steady state is dominated by quantum activation.}
 \label{tunnell}
\end{figure}

Now, the probability density at a given $g$ for times short compared to interwell processes but long compared to local relaxation is $\sim {\rm e}^{-R(g)/\Lambda}$, where $R(g)$ monotonously increases with $g$ from $R(g_{\rm min})=0$. 
Tunneling can now happen from the bottom of the well or at a higher quasi-energy $g$. If the product $e^{-2S_{tun}(g)/\Lambda} \, {\rm e}^{-R(g)/\Lambda}$ monotonously grows with $g$, this implies that the system has the tendency to tunnel towards increasing $g \to g_{\rm saddle}$, and eventually will go over the barrier.
%Thus, a monotonously with $g$ increasing product $e^{-2S_{tun}(g)/\Lambda} \, {\rm e}^{-R(g)/\Lambda}$  implies that the system has the tendency to tunnel towards increasing $g \to g_{\rm saddle}$, and eventually will go over the barrier. 
This phenomenon has been termed quantum activation. The condition for quantum activation is thus
\begin{equation}
2 \partial_g S_{\rm tun}(g)+\partial_g R(g)<0 \ \ \mbox{for}\ \ g_{\rm min}\leq g\leq g_{\rm saddle}\, .
\end{equation}
Since the derivative of the tunnel action provides the (dimensionless) tunneling time 
\begin{equation}
\tau(g)\equiv \partial_g S_{\rm tun}(g)= {\rm Im}\int \frac{dQ}{\partial_P g(Q,P)}\, ,
\end{equation}
the above condition can also be written as $\partial_g R(g)< 2|\tau(g)|$ with $\tau(g)<0$ in the respective energy range.  As Fig.~\ref{tunnell} reveals, this condition is always fulfilled so that we can conclude that the relaxation dynamics  towards a steady state  with rate $\Gamma_L$, as for example  seen in Fig.~\ref{Wignerlindz}, occurs dominantly via quantum activation. 

In essence, the picture that emerges from these and the previous sections is the following: Classically, the switching between period-3 states occurs via thermally activated processes, where a trajectory diffuses up in energy until it approaches the saddle points, from where it switches to another period-3 state via the period-1 oscillator as an intermediate state (for a detailed discussion see also \cite{zhangDykman2019}). Quantum mechanically, for sufficiently large friction, the system diffusively climbs up the quasi-energy ladder within a period-3 well until it approaches the energy range around the saddle point energy $g_{\rm saddle}$. In this energy range, direct tunneling through the quasi-energy barrier towards one of the adjacent period-3 states occurs.  The period-1 oscillator looses its role as an intermediate state since tunneling between period-3 and period-1 states is less likely for energetic reasons. However, it is still an open question how in detail the tunneling  between the period-3 states near the saddle points happens, a problem that we intend to consider elsewhere.

\section{Dissipative induced phase transition} 
\label{sec:transition}

	 \begin{figure*}[htb] 
\includegraphics[width=1\linewidth]{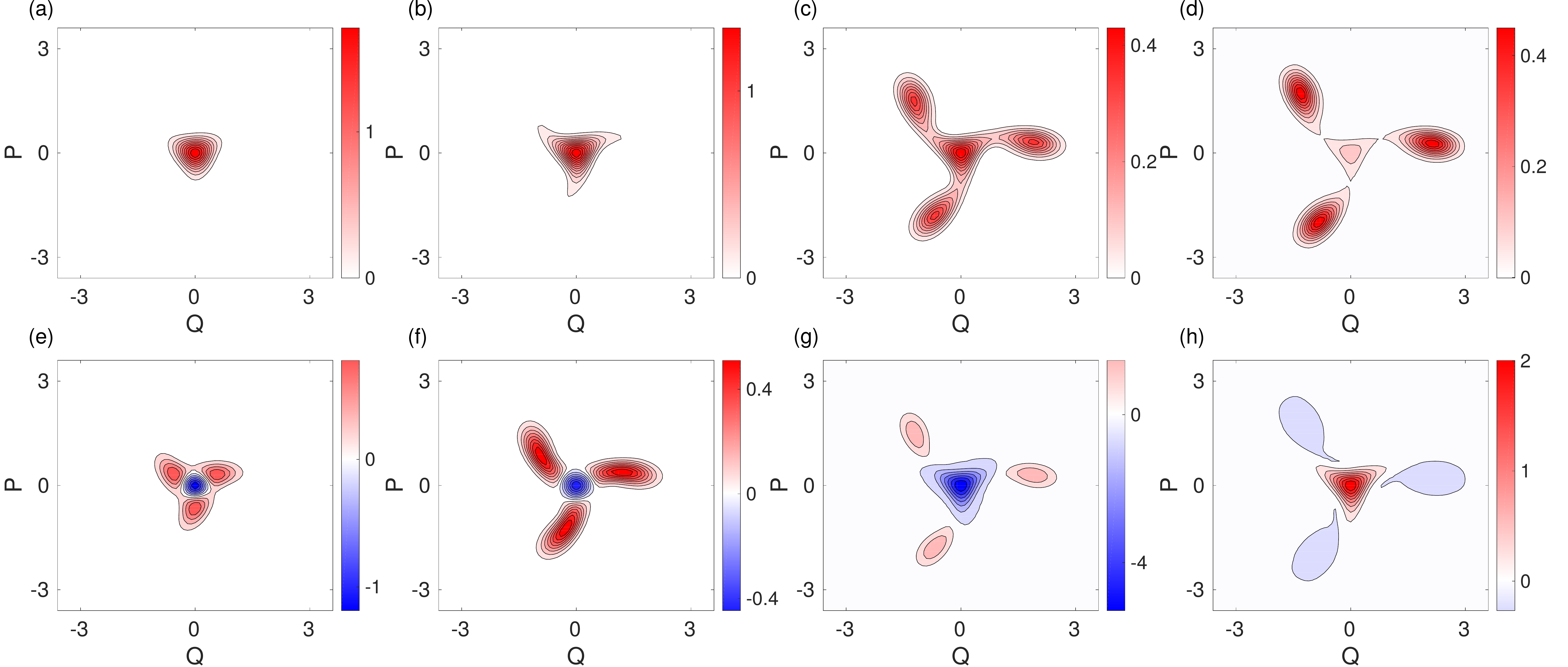}
 \caption{(a)-(d): Steady state Wigner density $\rho_W^{(0)}(Q,P)$ of $\ket{\ket{\nu_0}}$; (e)-(h): Wigner density $\rho_W^{(3,c)}(Q,P)$ of $\ket{\ket{\nu_3}}_{c}$ for various driving strengths $f=0.27$, $f=0.47$, $f=0.6$, $f=0.67$ (from (a) to (d) and (e) to (h)) at $\bar{n}=0$, $\kappa=1.5$ and $\Lambda=0.1688$, where $f_c\approx 0.63$ [cf.~Fig.~\ref{realimageigliou}(a)]. $\rho_W^{(0)}(Q,P)$ makes a transition from a localized state (ground state) around $Q=P=0$  for small driving to a delocalized distribution predominantly localized in the wells $Q_m, P_m, m=1, 2, 3$ corresponding to classical period-3 fixed points. $\rho_W^{(3,c)}(Q,P)$ exhibits the complementary behavior. }
 \label{wignerofphysexc}
\end{figure*}

 \begin{figure*}[htb] 
\includegraphics[width=1\textwidth]{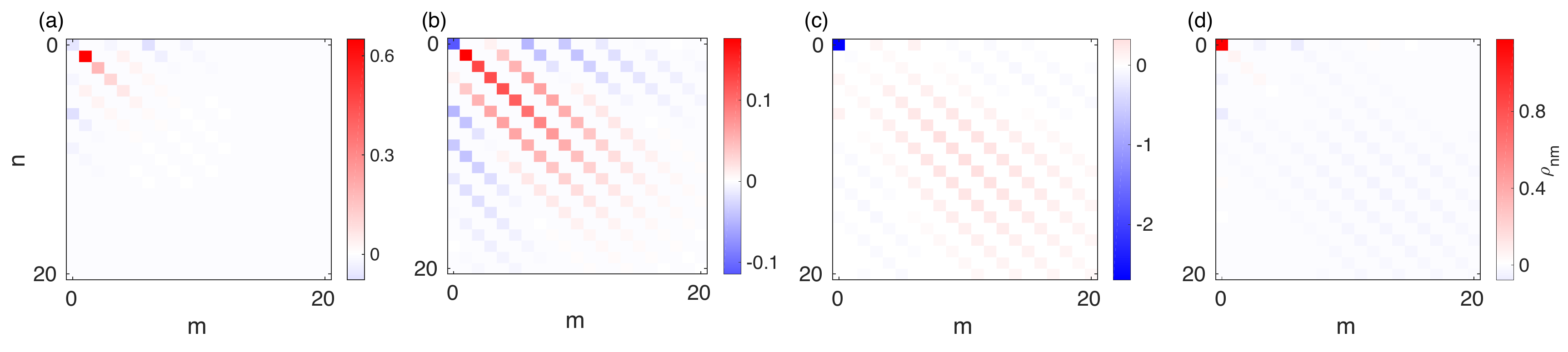}		
 \caption{Fock state representation of the steady state density $\langle n|\rho^{(3, c)}|m\rangle$ corresponding to the
 eigenvector $\ket{\ket{\nu_3}}_{c}$ orthogonal to the steady state for $f=0.27$, $f=0.47$, $f=0.6$, $f=0.67$ (from (a) to (d)) and $\kappa=1.5$, $\Lambda=0.1688$. The density has a dominant contribution in the first excited Fock state for small driving which turns into a dominant contribution in the ground state for strong driving.}
 \label{wignerofphysexch}
\end{figure*}

So far we have discussed the time scales for relaxation towards the steady state and its dependence on the driving strength. Now we turn to the steady state itself to explore to what extent the emergence of multiple steady state orbits at a critical driving in the classical regime yields signatures in the quantum realm. As we discussed in Sec.~\ref{escaperatelambda}, near the critical driving $f_c$, a new time scale appears that is associated with interwell processes. This coincides  with a fundamental change in the nature of the steady state from a localized to a delocalized one. We recall that for finite $\Lambda$ the transition is smeared out due to fluctuations.

To analyse this changeover in more detail, we come back to the eigenvectors $\ket{\ket{\nu_0}}$ and $\ket{\ket{\nu_3}}$ corresponding to the steady state density $\rho_{ss}\equiv \rho^{(0)}$ and to the eigen-density $\rho^{(3)}$, respectively. While the former is asymptotically approached for long times, the latter includes those populations in Fock space contained in the full time-dependent density that survive the longest. 
%carries those populations in Fock space of the full density that survive the longest. 
It turns out that the above transition leaves direct signatures in the Wigner and the Fock state representations of these two densities. As already mentioned above, the associated eigenvectors are not orthogonal. Consequently, in order to distill that part of $\ket{\ket{\nu_3}}$ that has no overlap with $\ket{\ket{\nu_0}}$, we consider 
\begin{equation}
\ket{\ket{\nu_3}}_c= N_3\left(\ket{\ket{\nu_3}}-\bra{\bra{\nu_0}}\ket{\ket{\nu_3}}\, \ket{\ket{\nu_0}}\right)\, 
\end{equation}
with normalization $N_3$. Of course, $\ket{\ket{\nu_3}}_c$ is also an eigenvector of $\mathcal{L}$ with eigenvalue $\lambda_3$.
Then, the corresponding density $\rho^{(3)}_c$ provides information about
those populations in Fock space, complementary to those in $\rho^{(0)}$,  that disappear with the least negative eigenvalue $\lambda_3$ during the relaxation process of the full density matrix $\rho(t)$ (the densities associated with $\lambda_{1, 2}$ carry off-diagonal elements only). One has to keep in mind though that it is not directly related to a physical density, but rather serves here as a diagnostic tool to analyse the (approximate) phase transition. 

Figure \ref{wignerofphysexc} illustrates how the corresponding Wigner densities evolve with increasing driving. While the steady state displays the expected transition from localized to delocalized (from strong overlap with the oscillator ground state at $Q=P=0$ to strong overlap with those around $Q_m, P_m; m>0$), the density $\rho^{(3)}_c$ shows the opposite behavior, from delocalized to localized. Accordingly, the period-1 oscillator at the origin turns from a stationary into a transient state, while period-3 oscillators are associated with steady states beyond the critical driving $f_c\approx 0.63$. 

The same behavior can be more precisely detected in the Fock state representation, see Fig. \ref{wignerofphysexch}. Below $f_c$ the distribution to $\rho^{(3)}_c$ occupies predominantly the {\em first excited} state of the oscillator at $Q=P=0$, i.e.\ $\rho^{(3)}_c\sim |1\rangle \langle1|$,  while beyond $f_c$ the {\em ground} state dominates, i.e.\ $\rho^{(3)}_c\sim |0\rangle \langle 0|$, and the excited state is absent. At that value of the driving, where this transition happens, i.e.\ $f_c$, we find empirically that  $|\lambda_3|$ takes its minimal value [see Fig.~\ref{realimageigliou}(a)]. This resembles behavior such as the closing of energy gaps and the emergence of slow modes that can be observed in phase transitions. For the ranges of parameters explored here, particularly small $\Lambda$, and $\kappa>1$, this coincides with a minimum in $|\lambda_3|$. 
This still holds for $\Lambda\sim O(1)$ and stronger friction. 
%For larger $\Lambda\sim O(1)$ friction has to increase.
 Outside of these domains, one observes a shift of  the critical driving towards smaller values (not shown).

	 \begin{figure}[htp]
\centering  
\includegraphics[width=1\linewidth]{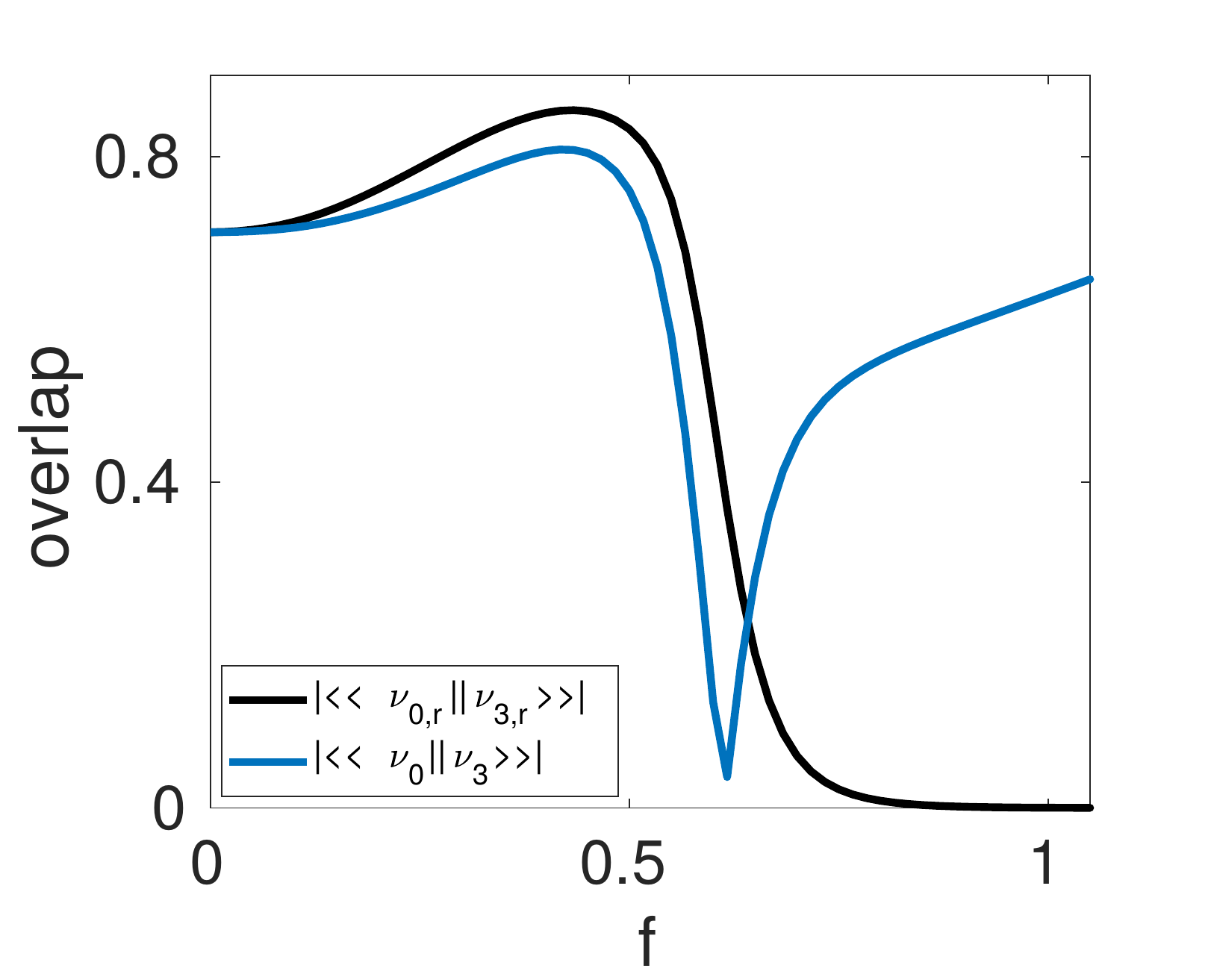}
 \caption{
The overlap $|\bra{\bra{\nu_0}}\ket{\ket{\nu_3}}|$ for $\kappa=1.5$, $\Lambda=0.1688$ (blue line) exhibits a maximum followed by a sharp drop to almost zero when $\ket{\ket{\nu_3}}$ becomes least unstable ($|\lambda_3|$ has a minimum); it sharply increases again with growing driving. Most of this behavior results from the components of the eigenvectors $\ket{\ket{\nu_0}}$ and $\ket{\ket{\nu_3}}$ in the sub-space spanned by the two lowest lying  Fock states $\{|0\rangle, |1\rangle\}$ ($|\bra{\bra{\nu_{0,r}}}\ket{\ket{\nu_{3,r}}}|$, black line).}
\label{realimageiglio}
\end{figure}

In terms of the original (non-orthogonal) eigenvectors, somewhat below the critical driving strength the overlap $|\bra{\bra{\nu_0}}\ket{\ket{\nu_3}}|$ (see Fig.~\ref{realimageiglio} blue line) exhibits a maximum followed by a sharp drop to almost zero when $\ket{\ket{\nu_3}}$ becomes least unstable ($|\lambda_3|$ has a minimum); it sharply increases again with growing driving. Most of this behavior results indeed from the components of the eigenvectors $\ket{\ket{\nu_0}}$ and $\ket{\ket{\nu_3}}$ in the sub-space spanned by the two lowest lying  Fock states $\{|0\rangle, |1\rangle\}$, i.e.\ $\ket{\ket{\nu_{0, r}}}$, $\ket{\ket{\nu_{3, r}}}$, see Fig.~\ref{realimageiglio} black line. 

\begin{figure}
\centering  
\includegraphics[width=1\linewidth]{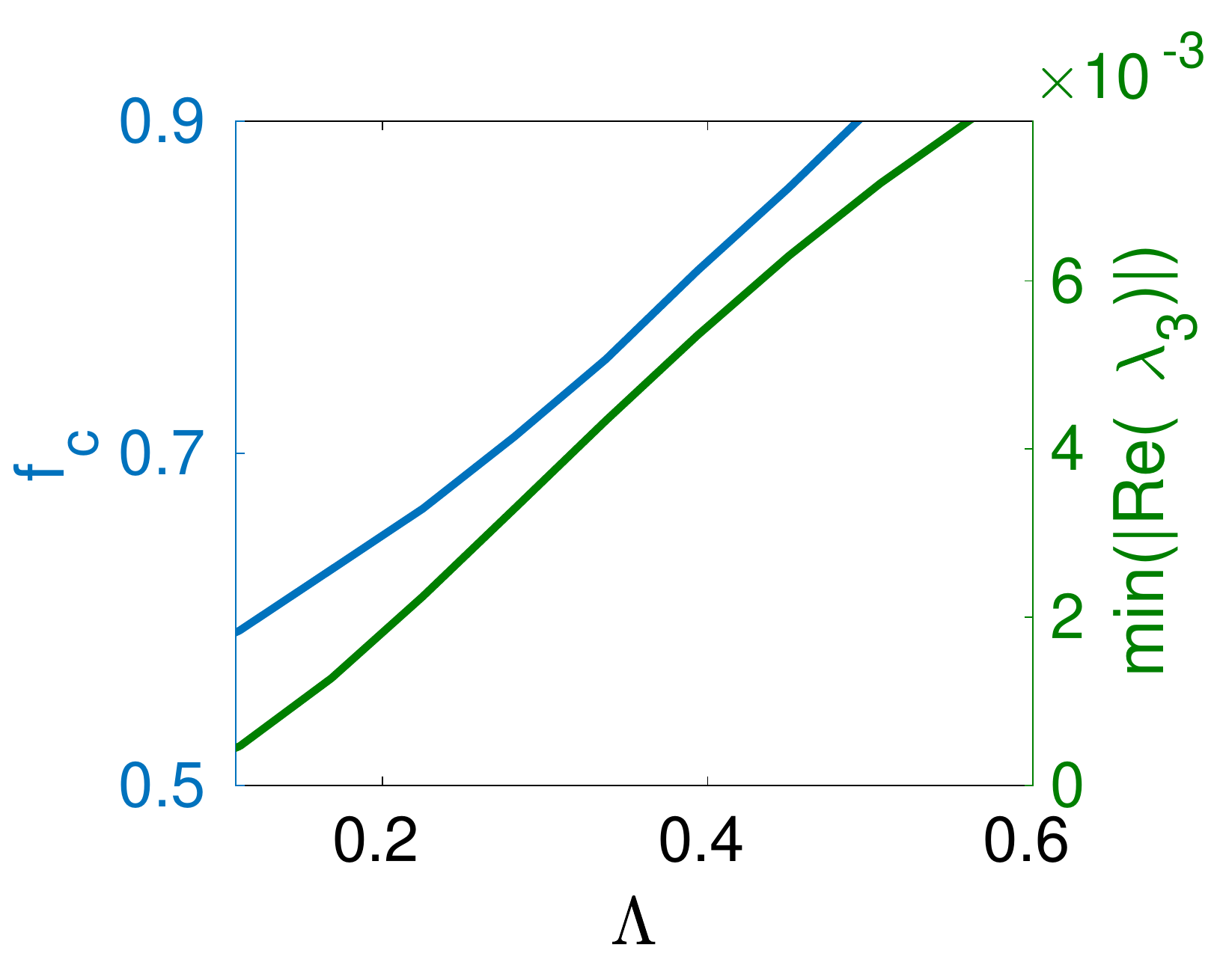}
 \caption{Critical driving $f_c$ (blue line) located at the minimum of the eigenvalue $|\lambda_3|$ with changing $\Lambda$ ($\kappa=1.5$ and $\bar{n}=0$). For decreasing $\Lambda$ the systems turns into the semiclassical regime and approaches the classical threshold $f_{c,cl}\approx 0.49$ for $\Lambda\to 0$. $\text{min}(|\rm Re(\lambda_3)|)$ (green line) decreases as $\Lambda$ goes to zero.}
 \label{changeinlambda}
\end{figure}

How the location of the minimum of $|\lambda_{3}|$ changes with the driving $f$ and the quantum scale $\Lambda$ is depicted in Fig.~\ref{changeinlambda}. For $\Lambda\to 0$, one  approaches the classical threshold $f_{c,cl}\approx 0.49$ (cf.~Fig.~\ref{bifdiag}), where the classical limit cannot be fully resolved (cf.~Fig.~\ref{changeinlambda}). With growing impact of quantum fluctuations $f_c(\Lambda)$ increases. Fig.~\ref{changeinlambda} also shows how $\text{min}(|\rm Re(\lambda_3)|)$ diminishes as the quantum fluctuations $\Lambda$ go to zero and the gap closes.

 \begin{figure}
\centering  
\includegraphics[width=1\linewidth]{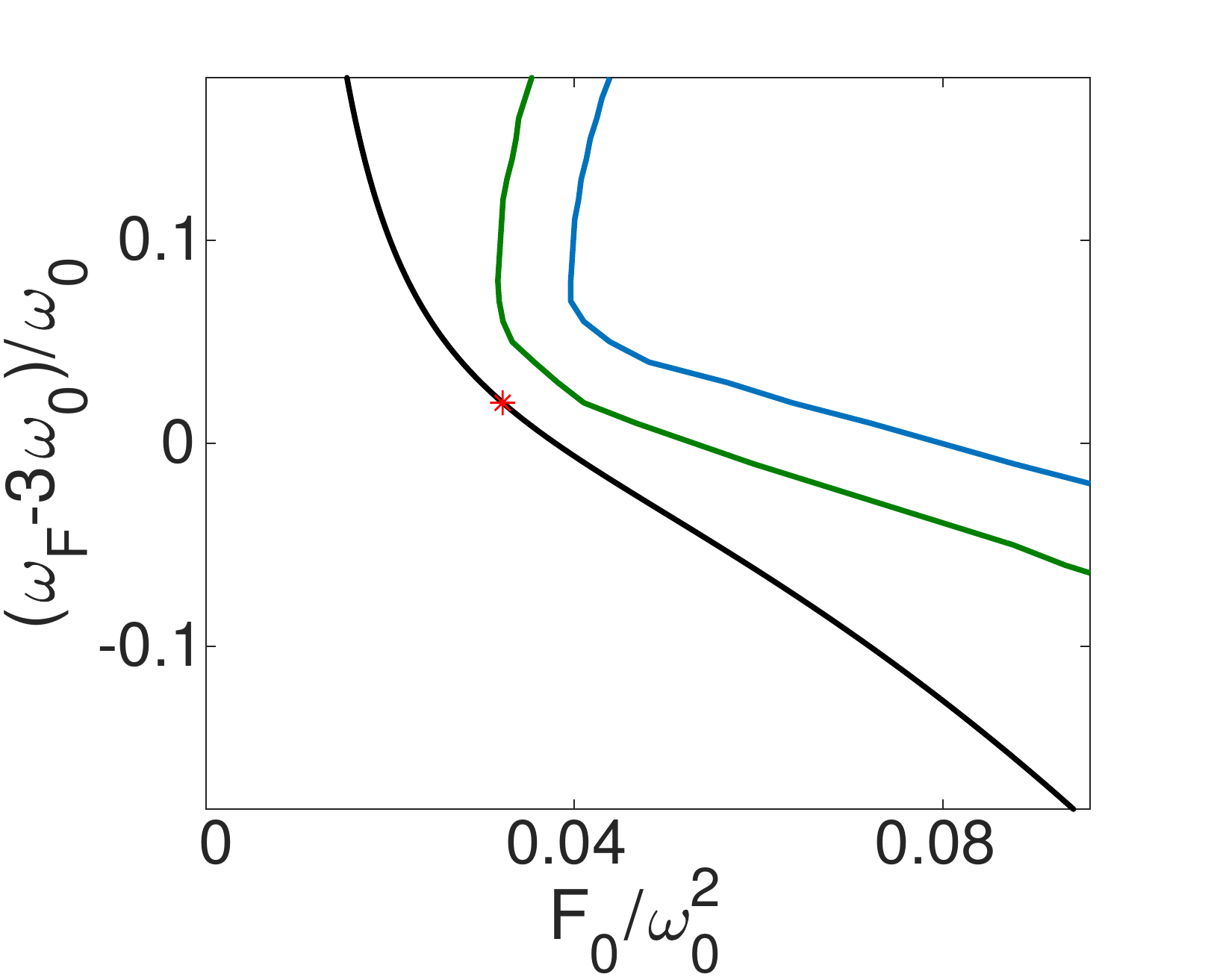}
 \caption{Phase diagram (de-tuning versus driving amplitude) of a dissipative quantum oscillator driven by three times its fundamental frequency. The black line separates the classical regions of single (left) and multiple (right) solutions. Quantum mechanically, the green ($\bar{n}=0$)/ blue ($\bar{n}=0.1$) line separates the domain, where the steady state is localized around the origin of the $QP$-plane (left) from the domain, where it is localized with dominant contributions around classical period-3 fixed points (right). In comparison to the classical prediction for the bifurcation line in absence of noise which separates the regime of period-1 orbits from the one of period-3 orbits, the quantum system exhibits ranges with a re-entrant behavior when increasing the de-tuning for fixed driving. Parameters are $\gamma/\omega_0=0.01$, $\alpha q_0^2/\omega_0^2=0.0015$ with the new quantum parameter $q_0^2=\hbar/(2M\omega_0)=1/2$. The red point corresponds to the transition point in Fig.~\ref{bifdiag} ($f_{c, \rm cl}= 0.49$).  }
 \label{quphaset}
\end{figure}
One unique feature of periodically driven oscillators lies in the fact that they allow exploring dissipative phase transitions in steady state within set-ups which are easily tunable by the parameters of the external drive. In our situation, the corresponding parameter space is the two-dimensional space $(\delta\omega, F_0)$. In order to characterize in this space the domains of localized and delocalized steady states, respectively, and thus to provide predictions for experimental realizations, one has to work with dimensionless quantities where the scaling does {\em not} include $\delta\omega$, as we have done so far. Accordingly, we here switch to another scaling and consider physical parameters. 
Figure \ref{quphaset} displays the corresponding phase diagram of the quantum system in the parameter space $([\omega_F-3\omega_0]/\omega_0, F_0/\omega_0^2)$. 
The green/ blue solid line separates for vanishing/ finite temperature the domain, where the steady state is localized (left area), from the domain, where it is delocalized with most of the weight sitting in the three well regions of the quasi-energy surface (right area). Apparently, tuning $[\omega_F-3\omega_0]/\omega_0$ for fixed driving $F_0/\omega_0^2$, one finds ranges for the driving, where a re-entrant behavior can be seen: one starts at a given $F_0/\omega_0^2$ for low de-tuning from a localized state that turns into  a delocalized one for growing $[\omega_F-3\omega_0]/\omega_0$, but ends again in a localized state for large de-tunings. In comparison we show the classical situation at $\text{T}=0$, where the re-entrant property is absent (black line).

\section{Conclusion}

In this work we studied quantum-noise induced switching between stable states of an oscillator that displays stationary period-three vibrations. The oscillator's quasi-energy consists of three localized wells in phase space when a sufficiently strong driving is applied. Starting initially from a locally relaxed state in one of the wells in phase space, transitions to the other wells occur via quantum activation rather than through quantum tunneling. The relaxation dynamics is characterized by a single time scale that separates interwell from intrawell processes. A standard semiclassical treatment does not fully capture the complexity of the dynamics and, thus, can at best yield the order of magnitude for this time scale. It also reveals a breaking of detailed balance for all temperatures down to $\text{T}=0$. 
%This time scale cannot be derived from a semiclassical treatment. The latter also reveals a breaking of detailed balance for all temperatures down to $T=0$. 
The appearance of a time scale separation is associated with an approximate closing of the gap between the lowest lying eigenvalues for the relaxation of populations and in turn a dissipation-induced transition in the nature of the stationary state from being localized to being delocalized. In contrast to the classical situation, this phase transition exhibits a re-entrant behavior in the parameter space of the external drive (amplitude, frequency). These findings may further stimulate on-going experimental investigations based on superconducting circuits.

\section*{acknowledgments}
We are grateful for financial support by the German Science Foundation through DFG AN336/11-1 and the Center for Integrated Quantum Science and Technology (IQST). We acknowledge fruitful discussions with M. I. Dykman and Y. Zhang.

\clearpage
\appendix

\section{Classical nonlinear oscillator} 
\label{appendix}

\subsection{Bifurcation diagram}

We start with  $q=\tilde{A}e^{\frac{i \omega_F t}{3}}+ \tilde{A}^* e^{-\frac{i \omega_F t}{3}} = A \cos(\frac{\omega_F}{3} t + \varphi)$ for a transformation of equation \eqref{eqofmot} to a rotating frame with $\omega_F \approx 3 \omega_0$ and $\tilde{A}=\tilde{A}'+i \tilde{A}''=A e^{i \varphi}$. This leads to the following equation
\begin{equation}
\tilde{A} = \frac{F_0}{2 i \omega_0} \tilde{A}^{*2} \frac{1}{(-\frac{\delta \omega}{i} +2 \gamma + \frac{3 \alpha}{i \omega_0}|\tilde{A}|^2)}  
\label{biftwoeq}
\end{equation}
that provides the amplitude as function of the three parameters $F_0, \alpha, \delta\omega$. Scaling the amplitude $A= 2 \sqrt{|\tilde{A}|^2}$ with the parameter $C=\sqrt{8\omega_0\delta\omega/(3\alpha)}$ and $F_0= f 3\sqrt{24 \omega_0 \alpha \delta \omega}$ as in the main text, one obtains the dimensionless amplitudes
\begin{equation}
\frac{A^2}{C^2}= \frac{9}{8} f^2 +\frac{1}{8} \pm \frac{1}{8}\left[(9 f^2+1)^2-4 \kappa^2-1\right]^{1/2}
\end{equation}
with $\kappa=\gamma/\delta\omega$. Multiple orbits coexist and the expression under the square root is positive for driving $f>f_{c, \rm cl}$ with
\begin{equation}
f_{c, \rm cl}=\frac{1}{3} \left(\sqrt{4\kappa^2+1}-1\right)^{1/2}\, .
\end{equation}

\subsection{Stability analysis}

For a stability analysis a small deviation $\delta \tilde{A}$ is included for the amplitude
\begin{equation}
\tilde{A} = \tilde{A}_{st} + \delta \tilde{A}, \hspace{0.5cm}
\tilde{A}^*=\tilde{A}_{st}^* + \delta \tilde{A}^*
\end{equation}
in equation \eqref{biftwoeq}. The stability is determined by the eigenvalues of the matrix of the set of equations
\begin{equation}
\begin{pmatrix}
\delta \dot{\tilde{A}} \\
\delta \dot{\tilde{A}}^*
\end{pmatrix}= 
\begin{pmatrix}
\frac{\delta \omega}{2 i} - \gamma - \frac{3 \alpha}{ i \omega_0} |\tilde{A}|^2 & \frac{F_0}{2 i \omega_0} \tilde{A}_{st}^*- \frac{3 \alpha}{2 i \omega_0} \tilde{A}_{st}^2  \\
-\frac{F_0}{2 i \omega_0} \tilde{A}_{st}+ \frac{3 \alpha}{2 i \omega_0} \tilde{A}_{st}^{*2} & -\frac{\delta \omega}{2 i} - \gamma + \frac{3 \alpha}{ i \omega_0} |\tilde{A}|^2
\end{pmatrix}
\begin{pmatrix}
\delta \tilde{A}\\
\delta \tilde{A}^*
\end{pmatrix}.
\end{equation}

\subsection{Experimental realisation} 
Classically we investigate numerically different types of driving to find a transition between period-1 and period-3 solutions, that could possibly be realised experimentally. One possibility uses a small linear drive until the system equilibrates at time $t_{equ}$ and then turns on a parametric drive to achieve a transition. The equation of motion results as
% One possibility would be to use a small linear drive until the system equilibrates at time $t_{equ}$ and then turn on a parametric drive to achieve a transition. The equation of motion would be
\begin{equation}
\ddot{q} + 2 \gamma \dot{q} + \omega_0^2 q + \alpha q^3 = F_2 \cos(\omega_{F,2} t) + F q^2 \cos(\omega_{F} t)
\end{equation}
with $\omega_{F,2} =\frac{\omega_F}{3} \approx \omega_0$. Alternatively it is also possible to use $\omega_{F,2} =\omega_F \approx 3\omega_0$ when turning on $F$ at time $t_{equ}$. 

Another easier method is to use only a linear drive and $\omega_F \approx 3 \omega_0$ so that
\begin{equation}
\ddot{q} + 2 \gamma \dot{q} + \omega_0^2 q + \alpha q^3 = F_2 \cos(\omega_{F} t)\, .
\end{equation}
We  make an ansatz $q=q^{(0)} +q^{(1)}$ with the leading order (linear) solution
\begin{equation}
  q^{(0)}= \frac{F_2 \cos(\omega_{F} t)}{\omega_0^2-\omega_{F}^2} \approx -\frac{F_2 \cos(\omega_{F_2} t)}{8 \omega_0^2}\, .
\end{equation}
Plugging this solution into the above equation of motion and  keeping only leading terms yields Eq.~(\ref{eqofmot}) of the main text, i.e., 
\begin{equation}
\begin{split}
\ddot{q}^{(1)} + 2 \gamma \dot{q}^{(1)} + \omega_0^2 q^{(1)} + \alpha q^{(1)^3} &= -3 \alpha q^{(1)^2} q^{(0)} \\
&= F_0  q^{(1)^2}  \cos(\omega_{F}t)
\end{split}
\end{equation}
with the drive parameter
\begin{equation}
F_0=\frac{3 \alpha}{8 \omega_0^2} F_2\, .
\end{equation}

\bibliography{literature}

%merlin.mbs apsrev4-1.bst 2010-07-25 4.21a (PWD, AO, DPC) hacked
%Control: key (0)
%Control: author (72) initials jnrlst
%Control: editor formatted (1) identically to author
%Control: production of article title (-1) disabled
%Control: page (0) single
%Control: year (1) truncated
%Control: production of eprint (0) enabled
\begin{thebibliography}{48}%
\makeatletter
\providecommand \@ifxundefined [1]{%
 \@ifx{#1\undefined}
}%
\providecommand \@ifnum [1]{%
 \ifnum #1\expandafter \@firstoftwo
 \else \expandafter \@secondoftwo
 \fi
}%
\providecommand \@ifx [1]{%
 \ifx #1\expandafter \@firstoftwo
 \else \expandafter \@secondoftwo
 \fi
}%
\providecommand \natexlab [1]{#1}%
\providecommand \enquote  [1]{``#1''}%
\providecommand \bibnamefont  [1]{#1}%
\providecommand \bibfnamefont [1]{#1}%
\providecommand \citenamefont [1]{#1}%
\providecommand \href@noop [0]{\@secondoftwo}%
\providecommand \href [0]{\begingroup \@sanitize@url \@href}%
\providecommand \@href[1]{\@@startlink{#1}\@@href}%
\providecommand \@@href[1]{\endgroup#1\@@endlink}%
\providecommand \@sanitize@url [0]{\catcode `\\12\catcode `\$12\catcode
  `\&12\catcode `\#12\catcode `\^12\catcode `\_12\catcode `\%12\relax}%
\providecommand \@@startlink[1]{}%
\providecommand \@@endlink[0]{}%
\providecommand \url  [0]{\begingroup\@sanitize@url \@url }%
\providecommand \@url [1]{\endgroup\@href {#1}{\urlprefix }}%
\providecommand \urlprefix  [0]{URL }%
\providecommand \Eprint [0]{\href }%
\providecommand \doibase [0]{http://dx.doi.org/}%
\providecommand \selectlanguage [0]{\@gobble}%
\providecommand \bibinfo  [0]{\@secondoftwo}%
\providecommand \bibfield  [0]{\@secondoftwo}%
\providecommand \translation [1]{[#1]}%
\providecommand \BibitemOpen [0]{}%
\providecommand \bibitemStop [0]{}%
\providecommand \bibitemNoStop [0]{.\EOS\space}%
\providecommand \EOS [0]{\spacefactor3000\relax}%
\providecommand \BibitemShut  [1]{\csname bibitem#1\endcsname}%
\let\auto@bib@innerbib\@empty
%</preamble>
\bibitem [{\citenamefont {H\"anggi}\ \emph {et~al.}(1990)\citenamefont
  {H\"anggi}, \citenamefont {Talkner},\ and\ \citenamefont
  {Borkovec}}]{RevModPhys.62.251}%
  \BibitemOpen
  \bibfield  {author} {\bibinfo {author} {\bibfnamefont {P.}~\bibnamefont
  {H\"anggi}}, \bibinfo {author} {\bibfnamefont {P.}~\bibnamefont {Talkner}}, \
  and\ \bibinfo {author} {\bibfnamefont {M.}~\bibnamefont {Borkovec}},\ }\href
  {\doibase 10.1103/RevModPhys.62.251} {\bibfield  {journal} {\bibinfo
  {journal} {Rev. Mod. Phys.}\ }\textbf {\bibinfo {volume} {62}},\ \bibinfo
  {pages} {251} (\bibinfo {year} {1990})}\BibitemShut {NoStop}%
\bibitem [{\citenamefont {Weiss}(2012)}]{Weiss2012}%
  \BibitemOpen
  \bibfield  {author} {\bibinfo {author} {\bibfnamefont {U.}~\bibnamefont
  {Weiss}},\ }\href@noop {} {\emph {\bibinfo {title} {Quantum Dissipative
  Systems}}}\ (\bibinfo  {publisher} {World Scientific Publishing Company},\
  \bibinfo {year} {2012})\BibitemShut {NoStop}%
\bibitem [{\citenamefont {Else}\ \emph {et~al.}(2016)\citenamefont {Else},
  \citenamefont {Bauer},\ and\ \citenamefont {Nayak}}]{PhysRevLett.117.090402}%
  \BibitemOpen
  \bibfield  {author} {\bibinfo {author} {\bibfnamefont {D.~V.}\ \bibnamefont
  {Else}}, \bibinfo {author} {\bibfnamefont {B.}~\bibnamefont {Bauer}}, \ and\
  \bibinfo {author} {\bibfnamefont {C.}~\bibnamefont {Nayak}},\ }\href
  {\doibase 10.1103/PhysRevLett.117.090402} {\bibfield  {journal} {\bibinfo
  {journal} {Phys. Rev. Lett.}\ }\textbf {\bibinfo {volume} {117}},\ \bibinfo
  {pages} {090402} (\bibinfo {year} {2016})}\BibitemShut {NoStop}%
\bibitem [{\citenamefont {Wilczek}(2012)}]{PhysRevLett.109.160401}%
  \BibitemOpen
  \bibfield  {author} {\bibinfo {author} {\bibfnamefont {F.}~\bibnamefont
  {Wilczek}},\ }\href {\doibase 10.1103/PhysRevLett.109.160401} {\bibfield
  {journal} {\bibinfo  {journal} {Phys. Rev. Lett.}\ }\textbf {\bibinfo
  {volume} {109}},\ \bibinfo {pages} {160401} (\bibinfo {year}
  {2012})}\BibitemShut {NoStop}%
\bibitem [{\citenamefont {Watanabe}\ and\ \citenamefont
  {Oshikawa}(2015)}]{PhysRevLett.114.251603}%
  \BibitemOpen
  \bibfield  {author} {\bibinfo {author} {\bibfnamefont {H.}~\bibnamefont
  {Watanabe}}\ and\ \bibinfo {author} {\bibfnamefont {M.}~\bibnamefont
  {Oshikawa}},\ }\href {\doibase 10.1103/PhysRevLett.114.251603} {\bibfield
  {journal} {\bibinfo  {journal} {Phys. Rev. Lett.}\ }\textbf {\bibinfo
  {volume} {114}},\ \bibinfo {pages} {251603} (\bibinfo {year}
  {2015})}\BibitemShut {NoStop}%
\bibitem [{\citenamefont {Chandran}\ and\ \citenamefont
  {Sondhi}(2016)}]{PhysRevB.93.174305}%
  \BibitemOpen
  \bibfield  {author} {\bibinfo {author} {\bibfnamefont {A.}~\bibnamefont
  {Chandran}}\ and\ \bibinfo {author} {\bibfnamefont {S.~L.}\ \bibnamefont
  {Sondhi}},\ }\href {\doibase 10.1103/PhysRevB.93.174305} {\bibfield
  {journal} {\bibinfo  {journal} {Phys. Rev. B}\ }\textbf {\bibinfo {volume}
  {93}},\ \bibinfo {pages} {174305} (\bibinfo {year} {2016})}\BibitemShut
  {NoStop}%
\bibitem [{\citenamefont {Khemani}\ \emph {et~al.}(2016)\citenamefont
  {Khemani}, \citenamefont {Lazarides}, \citenamefont {Moessner},\ and\
  \citenamefont {Sondhi}}]{PhysRevLett.116.250401}%
  \BibitemOpen
  \bibfield  {author} {\bibinfo {author} {\bibfnamefont {V.}~\bibnamefont
  {Khemani}}, \bibinfo {author} {\bibfnamefont {A.}~\bibnamefont {Lazarides}},
  \bibinfo {author} {\bibfnamefont {R.}~\bibnamefont {Moessner}}, \ and\
  \bibinfo {author} {\bibfnamefont {S.~L.}\ \bibnamefont {Sondhi}},\ }\href
  {\doibase 10.1103/PhysRevLett.116.250401} {\bibfield  {journal} {\bibinfo
  {journal} {Phys. Rev. Lett.}\ }\textbf {\bibinfo {volume} {116}},\ \bibinfo
  {pages} {250401} (\bibinfo {year} {2016})}\BibitemShut {NoStop}%
\bibitem [{\citenamefont {von Keyserlingk}\ and\ \citenamefont
  {Sondhi}(2016)}]{PhysRevB.93.245146}%
  \BibitemOpen
  \bibfield  {author} {\bibinfo {author} {\bibfnamefont {C.~W.}\ \bibnamefont
  {von Keyserlingk}}\ and\ \bibinfo {author} {\bibfnamefont {S.~L.}\
  \bibnamefont {Sondhi}},\ }\href {\doibase 10.1103/PhysRevB.93.245146}
  {\bibfield  {journal} {\bibinfo  {journal} {Phys. Rev. B}\ }\textbf {\bibinfo
  {volume} {93}},\ \bibinfo {pages} {245146} (\bibinfo {year}
  {2016})}\BibitemShut {NoStop}%
\bibitem [{\citenamefont {Yao}\ \emph {et~al.}(2017)\citenamefont {Yao},
  \citenamefont {Potter}, \citenamefont {Potirniche},\ and\ \citenamefont
  {Vishwanath}}]{PhysRevLett.118.030401}%
  \BibitemOpen
  \bibfield  {author} {\bibinfo {author} {\bibfnamefont {N.~Y.}\ \bibnamefont
  {Yao}}, \bibinfo {author} {\bibfnamefont {A.~C.}\ \bibnamefont {Potter}},
  \bibinfo {author} {\bibfnamefont {I.-D.}\ \bibnamefont {Potirniche}}, \ and\
  \bibinfo {author} {\bibfnamefont {A.}~\bibnamefont {Vishwanath}},\ }\href
  {\doibase 10.1103/PhysRevLett.118.030401} {\bibfield  {journal} {\bibinfo
  {journal} {Phys. Rev. Lett.}\ }\textbf {\bibinfo {volume} {118}},\ \bibinfo
  {pages} {030401} (\bibinfo {year} {2017})}\BibitemShut {NoStop}%
\bibitem [{\citenamefont {Sacha}\ and\ \citenamefont
  {Zakrzewski}(2017)}]{Sacha_2017}%
  \BibitemOpen
  \bibfield  {author} {\bibinfo {author} {\bibfnamefont {K.}~\bibnamefont
  {Sacha}}\ and\ \bibinfo {author} {\bibfnamefont {J.}~\bibnamefont
  {Zakrzewski}},\ }\href {\doibase 10.1088/1361-6633/aa8b38} {\bibfield
  {journal} {\bibinfo  {journal} {Rep. Prog. Phys.}\ }\textbf {\bibinfo
  {volume} {81}},\ \bibinfo {pages} {016401} (\bibinfo {year}
  {2017})}\BibitemShut {NoStop}%
\bibitem [{\citenamefont {Kitagawa}\ \emph {et~al.}(2012)\citenamefont
  {Kitagawa}, \citenamefont {Broome}, \citenamefont {Fedrizzi}, \citenamefont
  {Rudner}, \citenamefont {Berg}, \citenamefont {Kassal}, \citenamefont
  {Aspuru-Guzik}, \citenamefont {Demler},\ and\ \citenamefont
  {White}}]{Kitagawa2012}%
  \BibitemOpen
  \bibfield  {author} {\bibinfo {author} {\bibfnamefont {T.}~\bibnamefont
  {Kitagawa}}, \bibinfo {author} {\bibfnamefont {M.~A.}\ \bibnamefont
  {Broome}}, \bibinfo {author} {\bibfnamefont {A.}~\bibnamefont {Fedrizzi}},
  \bibinfo {author} {\bibfnamefont {M.~S.}\ \bibnamefont {Rudner}}, \bibinfo
  {author} {\bibfnamefont {E.}~\bibnamefont {Berg}}, \bibinfo {author}
  {\bibfnamefont {I.}~\bibnamefont {Kassal}}, \bibinfo {author} {\bibfnamefont
  {A.}~\bibnamefont {Aspuru-Guzik}}, \bibinfo {author} {\bibfnamefont
  {E.}~\bibnamefont {Demler}}, \ and\ \bibinfo {author} {\bibfnamefont {A.~G.}\
  \bibnamefont {White}},\ }\href@noop {} {\bibfield  {journal} {\bibinfo
  {journal} {Nat. Commun.}\ }\textbf {\bibinfo {volume} {3}},\ \bibinfo {pages}
  {882} (\bibinfo {year} {2012})}\BibitemShut {NoStop}%
\bibitem [{\citenamefont {Jiang}\ \emph {et~al.}(2011)\citenamefont {Jiang},
  \citenamefont {Kitagawa}, \citenamefont {Alicea}, \citenamefont {Akhmerov},
  \citenamefont {Pekker}, \citenamefont {Refael}, \citenamefont {Cirac},
  \citenamefont {Demler}, \citenamefont {Lukin},\ and\ \citenamefont
  {Zoller}}]{PhysRevLett.106.220402}%
  \BibitemOpen
  \bibfield  {author} {\bibinfo {author} {\bibfnamefont {L.}~\bibnamefont
  {Jiang}}, \bibinfo {author} {\bibfnamefont {T.}~\bibnamefont {Kitagawa}},
  \bibinfo {author} {\bibfnamefont {J.}~\bibnamefont {Alicea}}, \bibinfo
  {author} {\bibfnamefont {A.~R.}\ \bibnamefont {Akhmerov}}, \bibinfo {author}
  {\bibfnamefont {D.}~\bibnamefont {Pekker}}, \bibinfo {author} {\bibfnamefont
  {G.}~\bibnamefont {Refael}}, \bibinfo {author} {\bibfnamefont {J.~I.}\
  \bibnamefont {Cirac}}, \bibinfo {author} {\bibfnamefont {E.}~\bibnamefont
  {Demler}}, \bibinfo {author} {\bibfnamefont {M.~D.}\ \bibnamefont {Lukin}}, \
  and\ \bibinfo {author} {\bibfnamefont {P.}~\bibnamefont {Zoller}},\ }\href
  {\doibase 10.1103/PhysRevLett.106.220402} {\bibfield  {journal} {\bibinfo
  {journal} {Phys. Rev. Lett.}\ }\textbf {\bibinfo {volume} {106}},\ \bibinfo
  {pages} {220402} (\bibinfo {year} {2011})}\BibitemShut {NoStop}%
\bibitem [{\citenamefont {Zhang}\ \emph
  {et~al.}(2017{\natexlab{a}})\citenamefont {Zhang}, \citenamefont {Hess},
  \citenamefont {Kyprianidis}, \citenamefont {Becker}, \citenamefont {Lee},
  \citenamefont {Smith}, \citenamefont {Pagano}, \citenamefont {Potirniche},
  \citenamefont {Potter}, \citenamefont {Vishwanath}, \citenamefont {Yao},\
  and\ \citenamefont {Monroe}}]{10.1038/nature21413}%
  \BibitemOpen
  \bibfield  {author} {\bibinfo {author} {\bibfnamefont {J.}~\bibnamefont
  {Zhang}}, \bibinfo {author} {\bibfnamefont {P.~W.}\ \bibnamefont {Hess}},
  \bibinfo {author} {\bibfnamefont {A.}~\bibnamefont {Kyprianidis}}, \bibinfo
  {author} {\bibfnamefont {P.}~\bibnamefont {Becker}}, \bibinfo {author}
  {\bibfnamefont {A.}~\bibnamefont {Lee}}, \bibinfo {author} {\bibfnamefont
  {J.}~\bibnamefont {Smith}}, \bibinfo {author} {\bibfnamefont
  {G.}~\bibnamefont {Pagano}}, \bibinfo {author} {\bibfnamefont {I.-D.}\
  \bibnamefont {Potirniche}}, \bibinfo {author} {\bibfnamefont {A.~C.}\
  \bibnamefont {Potter}}, \bibinfo {author} {\bibfnamefont {A.}~\bibnamefont
  {Vishwanath}}, \bibinfo {author} {\bibfnamefont {N.~Y.}\ \bibnamefont {Yao}},
  \ and\ \bibinfo {author} {\bibfnamefont {C.}~\bibnamefont {Monroe}},\
  }\href@noop {} {\bibfield  {journal} {\bibinfo  {journal} {Nature}\ }\textbf
  {\bibinfo {volume} {543}},\ \bibinfo {pages} {217} (\bibinfo {year}
  {2017}{\natexlab{a}})}\BibitemShut {NoStop}%
\bibitem [{\citenamefont {Choi}\ \emph {et~al.}(2017)\citenamefont {Choi},
  \citenamefont {Choi}, \citenamefont {Landig}, \citenamefont {Kucsko},
  \citenamefont {Zhou}, \citenamefont {Isoya}, \citenamefont {Jelezko},
  \citenamefont {Onoda}, \citenamefont {Sumiya}, \citenamefont {Khemani},
  \citenamefont {von Keyserlingk}, \citenamefont {Yao}, \citenamefont
  {Demler},\ and\ \citenamefont {Lukin}}]{nature21426}%
  \BibitemOpen
  \bibfield  {author} {\bibinfo {author} {\bibfnamefont {S.}~\bibnamefont
  {Choi}}, \bibinfo {author} {\bibfnamefont {J.}~\bibnamefont {Choi}}, \bibinfo
  {author} {\bibfnamefont {R.}~\bibnamefont {Landig}}, \bibinfo {author}
  {\bibfnamefont {G.}~\bibnamefont {Kucsko}}, \bibinfo {author} {\bibfnamefont
  {H.}~\bibnamefont {Zhou}}, \bibinfo {author} {\bibfnamefont {J.}~\bibnamefont
  {Isoya}}, \bibinfo {author} {\bibfnamefont {F.}~\bibnamefont {Jelezko}},
  \bibinfo {author} {\bibfnamefont {S.}~\bibnamefont {Onoda}}, \bibinfo
  {author} {\bibfnamefont {H.}~\bibnamefont {Sumiya}}, \bibinfo {author}
  {\bibfnamefont {V.}~\bibnamefont {Khemani}}, \bibinfo {author} {\bibfnamefont
  {C.}~\bibnamefont {von Keyserlingk}}, \bibinfo {author} {\bibfnamefont
  {N.~Y.}\ \bibnamefont {Yao}}, \bibinfo {author} {\bibfnamefont
  {E.}~\bibnamefont {Demler}}, \ and\ \bibinfo {author} {\bibfnamefont {M.~D.}\
  \bibnamefont {Lukin}},\ }\href@noop {} {\bibfield  {journal} {\bibinfo
  {journal} {Nature}\ }\textbf {\bibinfo {volume} {543}},\ \bibinfo {pages}
  {221} (\bibinfo {year} {2017})}\BibitemShut {NoStop}%
\bibitem [{\citenamefont {M.Dykman}(2012)}]{Dykmanbook2012}%
  \BibitemOpen
  \bibfield  {author} {\bibinfo {author} {\bibnamefont {M.Dykman}},\
  }\href@noop {} {\emph {\bibinfo {title} {Fluctuating Nonlinear Oscillators:
  From Nanomechanics to Quantum Superconducting Circuits}}}\ (\bibinfo
  {publisher} {Oxford University Press},\ \bibinfo {year} {2012})\BibitemShut
  {NoStop}%
\bibitem [{\citenamefont {Poot}\ and\ \citenamefont {van~der
  Zant}(2012)}]{PootvanderZant2012}%
  \BibitemOpen
  \bibfield  {author} {\bibinfo {author} {\bibfnamefont {M.}~\bibnamefont
  {Poot}}\ and\ \bibinfo {author} {\bibfnamefont {H.~S.}\ \bibnamefont {van~der
  Zant}},\ }\href {\doibase https://doi.org/10.1016/j.physrep.2011.12.004}
  {\bibfield  {journal} {\bibinfo  {journal} {Phys. Rep.}\ }\textbf {\bibinfo
  {volume} {511}},\ \bibinfo {pages} {273 } (\bibinfo {year}
  {2012})}\BibitemShut {NoStop}%
\bibitem [{\citenamefont {Dykman}\ and\ \citenamefont
  {Krivoglaz}(1980)}]{DykmanKrivoglaz1980}%
  \BibitemOpen
  \bibfield  {author} {\bibinfo {author} {\bibfnamefont {M.}~\bibnamefont
  {Dykman}}\ and\ \bibinfo {author} {\bibfnamefont {M.}~\bibnamefont
  {Krivoglaz}},\ }\href
  {http://www.sciencedirect.com/science/article/pii/0378437180900102}
  {\bibfield  {journal} {\bibinfo  {journal} {Physica A}\ }\textbf {\bibinfo
  {volume} {104}},\ \bibinfo {pages} {480 } (\bibinfo {year}
  {1980})}\BibitemShut {NoStop}%
\bibitem [{\citenamefont {Peano}\ and\ \citenamefont
  {Thorwart}(2006)}]{PeanoThorwart20062}%
  \BibitemOpen
  \bibfield  {author} {\bibinfo {author} {\bibfnamefont {V.}~\bibnamefont
  {Peano}}\ and\ \bibinfo {author} {\bibfnamefont {M.}~\bibnamefont
  {Thorwart}},\ }\href
  {http://www.sciencedirect.com/science/article/pii/S0301010405002636}
  {\bibfield  {journal} {\bibinfo  {journal} {Chem. Phys.}\ }\textbf {\bibinfo
  {volume} {322}},\ \bibinfo {pages} {135} (\bibinfo {year}
  {2006})}\BibitemShut {NoStop}%
\bibitem [{\citenamefont {Serban}\ and\ \citenamefont
  {Wilhelm}(2007)}]{SerbanWilhelm2007}%
  \BibitemOpen
  \bibfield  {author} {\bibinfo {author} {\bibfnamefont {I.}~\bibnamefont
  {Serban}}\ and\ \bibinfo {author} {\bibfnamefont {F.~K.}\ \bibnamefont
  {Wilhelm}},\ }\href {https://link.aps.org/doi/10.1103/PhysRevLett.99.137001}
  {\bibfield  {journal} {\bibinfo  {journal} {Phys. Rev. Lett.}\ }\textbf
  {\bibinfo {volume} {99}},\ \bibinfo {pages} {137001} (\bibinfo {year}
  {2007})}\BibitemShut {NoStop}%
\bibitem [{\citenamefont {Guo}\ \emph {et~al.}(2010)\citenamefont {Guo},
  \citenamefont {Zheng},\ and\ \citenamefont {Li}}]{GuoZhengLi2010}%
  \BibitemOpen
  \bibfield  {author} {\bibinfo {author} {\bibfnamefont {L.-Z.}\ \bibnamefont
  {Guo}}, \bibinfo {author} {\bibfnamefont {Z.-G.}\ \bibnamefont {Zheng}}, \
  and\ \bibinfo {author} {\bibfnamefont {X.-Q.}\ \bibnamefont {Li}},\ }\href
  {http://stacks.iop.org/0295-5075/90/i=1/a=10011} {\bibfield  {journal}
  {\bibinfo  {journal} {EPL}\ }\textbf {\bibinfo {volume} {90}},\ \bibinfo
  {pages} {10011} (\bibinfo {year} {2010})}\BibitemShut {NoStop}%
\bibitem [{\citenamefont {Guo}\ \emph {et~al.}(2011)\citenamefont {Guo},
  \citenamefont {Zheng}, \citenamefont {Li},\ and\ \citenamefont
  {Yan}}]{GuoZhengLiZhengLiYan2011}%
  \BibitemOpen
  \bibfield  {author} {\bibinfo {author} {\bibfnamefont {L.}~\bibnamefont
  {Guo}}, \bibinfo {author} {\bibfnamefont {Z.}~\bibnamefont {Zheng}}, \bibinfo
  {author} {\bibfnamefont {X.-Q.}\ \bibnamefont {Li}}, \ and\ \bibinfo {author}
  {\bibfnamefont {Y.~J.}\ \bibnamefont {Yan}},\ }\href {\doibase
  10.1103/PhysRevE.84.011144} {\bibfield  {journal} {\bibinfo  {journal} {Phys.
  Rev. E}\ }\textbf {\bibinfo {volume} {84}},\ \bibinfo {pages} {011144}
  (\bibinfo {year} {2011})}\BibitemShut {NoStop}%
\bibitem [{\citenamefont {Andr\'e}\ \emph {et~al.}(2012)\citenamefont
  {Andr\'e}, \citenamefont {Guo}, \citenamefont {Peano}, \citenamefont
  {Marthaler},\ and\ \citenamefont {Sch\"on}}]{andreguopeanoschoen2012}%
  \BibitemOpen
  \bibfield  {author} {\bibinfo {author} {\bibfnamefont {S.}~\bibnamefont
  {Andr\'e}}, \bibinfo {author} {\bibfnamefont {L.}~\bibnamefont {Guo}},
  \bibinfo {author} {\bibfnamefont {V.}~\bibnamefont {Peano}}, \bibinfo
  {author} {\bibfnamefont {M.}~\bibnamefont {Marthaler}}, \ and\ \bibinfo
  {author} {\bibfnamefont {G.}~\bibnamefont {Sch\"on}},\ }\href
  {https://link.aps.org/doi/10.1103/PhysRevA.85.053825} {\bibfield  {journal}
  {\bibinfo  {journal} {Phys. Rev. A}\ }\textbf {\bibinfo {volume} {85}},\
  \bibinfo {pages} {053825} (\bibinfo {year} {2012})}\BibitemShut {NoStop}%
\bibitem [{\citenamefont {Dykman}\ \emph {et~al.}(1998)\citenamefont {Dykman},
  \citenamefont {Maloney}, \citenamefont {Smelyanskiy},\ and\ \citenamefont
  {Silverstein}}]{DykmanMaloneySilverstein1998}%
  \BibitemOpen
  \bibfield  {author} {\bibinfo {author} {\bibfnamefont {M.~I.}\ \bibnamefont
  {Dykman}}, \bibinfo {author} {\bibfnamefont {C.~M.}\ \bibnamefont {Maloney}},
  \bibinfo {author} {\bibfnamefont {V.~N.}\ \bibnamefont {Smelyanskiy}}, \ and\
  \bibinfo {author} {\bibfnamefont {M.}~\bibnamefont {Silverstein}},\ }\href
  {\doibase 10.1103/PhysRevE.57.5202} {\bibfield  {journal} {\bibinfo
  {journal} {Phys. Rev. E}\ }\textbf {\bibinfo {volume} {57}},\ \bibinfo
  {pages} {5202} (\bibinfo {year} {1998})}\BibitemShut {NoStop}%
\bibitem [{\citenamefont {Zorin}\ and\ \citenamefont
  {Makhlin}(2011)}]{PhysRevB.83.224506}%
  \BibitemOpen
  \bibfield  {author} {\bibinfo {author} {\bibfnamefont {A.~B.}\ \bibnamefont
  {Zorin}}\ and\ \bibinfo {author} {\bibfnamefont {Y.}~\bibnamefont
  {Makhlin}},\ }\href {\doibase 10.1103/PhysRevB.83.224506} {\bibfield
  {journal} {\bibinfo  {journal} {Phys. Rev. B}\ }\textbf {\bibinfo {volume}
  {83}},\ \bibinfo {pages} {224506} (\bibinfo {year} {2011})}\BibitemShut
  {NoStop}%
\bibitem [{\citenamefont {Dykman}\ \emph {et~al.}(2011)\citenamefont {Dykman},
  \citenamefont {Marthaler},\ and\ \citenamefont
  {Peano}}]{DykmanMarthalerPeano2011}%
  \BibitemOpen
  \bibfield  {author} {\bibinfo {author} {\bibfnamefont {M.~I.}\ \bibnamefont
  {Dykman}}, \bibinfo {author} {\bibfnamefont {M.}~\bibnamefont {Marthaler}}, \
  and\ \bibinfo {author} {\bibfnamefont {V.}~\bibnamefont {Peano}},\ }\href
  {\doibase 10.1103/PhysRevA.83.052115} {\bibfield  {journal} {\bibinfo
  {journal} {Phys. Rev. A}\ }\textbf {\bibinfo {volume} {83}},\ \bibinfo
  {pages} {052115} (\bibinfo {year} {2011})}\BibitemShut {NoStop}%
\bibitem [{\citenamefont {Wustmann}\ and\ \citenamefont
  {Shumeiko}(2013)}]{PhysRevB.87.184501}%
  \BibitemOpen
  \bibfield  {author} {\bibinfo {author} {\bibfnamefont {W.}~\bibnamefont
  {Wustmann}}\ and\ \bibinfo {author} {\bibfnamefont {V.}~\bibnamefont
  {Shumeiko}},\ }\href {\doibase 10.1103/PhysRevB.87.184501} {\bibfield
  {journal} {\bibinfo  {journal} {Phys. Rev. B}\ }\textbf {\bibinfo {volume}
  {87}},\ \bibinfo {pages} {184501} (\bibinfo {year} {2013})}\BibitemShut
  {NoStop}%
\bibitem [{\citenamefont {Wustmann}\ and\ \citenamefont
  {Shumeiko}(2019)}]{doi10106315116533}%
  \BibitemOpen
  \bibfield  {author} {\bibinfo {author} {\bibfnamefont {W.}~\bibnamefont
  {Wustmann}}\ and\ \bibinfo {author} {\bibfnamefont {V.}~\bibnamefont
  {Shumeiko}},\ }\href {\doibase 10.1063/1.5116533} {\bibfield  {journal}
  {\bibinfo  {journal} {Low Tem. Phys.}\ }\textbf {\bibinfo {volume} {45}},\
  \bibinfo {pages} {848} (\bibinfo {year} {2019})}\BibitemShut {NoStop}%
\bibitem [{\citenamefont {Wustmann}\ and\ \citenamefont
  {Shumeiko}(2017)}]{PhysRevApplied.8.024018}%
  \BibitemOpen
  \bibfield  {author} {\bibinfo {author} {\bibfnamefont {W.}~\bibnamefont
  {Wustmann}}\ and\ \bibinfo {author} {\bibfnamefont {V.}~\bibnamefont
  {Shumeiko}},\ }\href {\doibase 10.1103/PhysRevApplied.8.024018} {\bibfield
  {journal} {\bibinfo  {journal} {Phys. Rev. Applied}\ }\textbf {\bibinfo
  {volume} {8}},\ \bibinfo {pages} {024018} (\bibinfo {year}
  {2017})}\BibitemShut {NoStop}%
\bibitem [{\citenamefont {Marthaler}\ and\ \citenamefont
  {Dykman}(2007)}]{PhysRevA.76.010102}%
  \BibitemOpen
  \bibfield  {author} {\bibinfo {author} {\bibfnamefont {M.}~\bibnamefont
  {Marthaler}}\ and\ \bibinfo {author} {\bibfnamefont {M.~I.}\ \bibnamefont
  {Dykman}},\ }\href {\doibase 10.1103/PhysRevA.76.010102} {\bibfield
  {journal} {\bibinfo  {journal} {Phys. Rev. A}\ }\textbf {\bibinfo {volume}
  {76}},\ \bibinfo {pages} {010102(R)} (\bibinfo {year} {2007})}\BibitemShut
  {NoStop}%
\bibitem [{\citenamefont {Peano}\ \emph {et~al.}(2012)\citenamefont {Peano},
  \citenamefont {Marthaler},\ and\ \citenamefont
  {Dykman}}]{PhysRevLett.109.090401}%
  \BibitemOpen
  \bibfield  {author} {\bibinfo {author} {\bibfnamefont {V.}~\bibnamefont
  {Peano}}, \bibinfo {author} {\bibfnamefont {M.}~\bibnamefont {Marthaler}}, \
  and\ \bibinfo {author} {\bibfnamefont {M.~I.}\ \bibnamefont {Dykman}},\
  }\href {\doibase 10.1103/PhysRevLett.109.090401} {\bibfield  {journal}
  {\bibinfo  {journal} {Phys. Rev. Lett.}\ }\textbf {\bibinfo {volume} {109}},\
  \bibinfo {pages} {090401} (\bibinfo {year} {2012})}\BibitemShut {NoStop}%
\bibitem [{\citenamefont {Marthaler}\ and\ \citenamefont
  {Dykman}(2006)}]{MarthalerDykmanswitchin2006}%
  \BibitemOpen
  \bibfield  {author} {\bibinfo {author} {\bibfnamefont {M.}~\bibnamefont
  {Marthaler}}\ and\ \bibinfo {author} {\bibfnamefont {M.~I.}\ \bibnamefont
  {Dykman}},\ }\href {\doibase 10.1103/PhysRevA.73.042108} {\bibfield
  {journal} {\bibinfo  {journal} {Phys. Rev. A}\ }\textbf {\bibinfo {volume}
  {73}},\ \bibinfo {pages} {042108} (\bibinfo {year} {2006})}\BibitemShut
  {NoStop}%
\bibitem [{\citenamefont {Ong}\ \emph {et~al.}(2013)\citenamefont {Ong},
  \citenamefont {Boissonneault}, \citenamefont {Mallet}, \citenamefont
  {Doherty}, \citenamefont {Blais}, \citenamefont {Vion}, \citenamefont
  {Esteve},\ and\ \citenamefont {Bertet}}]{PhysRevLett.110.047001}%
  \BibitemOpen
  \bibfield  {author} {\bibinfo {author} {\bibfnamefont {F.~R.}\ \bibnamefont
  {Ong}}, \bibinfo {author} {\bibfnamefont {M.}~\bibnamefont {Boissonneault}},
  \bibinfo {author} {\bibfnamefont {F.}~\bibnamefont {Mallet}}, \bibinfo
  {author} {\bibfnamefont {A.~C.}\ \bibnamefont {Doherty}}, \bibinfo {author}
  {\bibfnamefont {A.}~\bibnamefont {Blais}}, \bibinfo {author} {\bibfnamefont
  {D.}~\bibnamefont {Vion}}, \bibinfo {author} {\bibfnamefont {D.}~\bibnamefont
  {Esteve}}, \ and\ \bibinfo {author} {\bibfnamefont {P.}~\bibnamefont
  {Bertet}},\ }\href {\doibase 10.1103/PhysRevLett.110.047001} {\bibfield
  {journal} {\bibinfo  {journal} {Phys. Rev. Lett.}\ }\textbf {\bibinfo
  {volume} {110}},\ \bibinfo {pages} {047001} (\bibinfo {year}
  {2013})}\BibitemShut {NoStop}%
\bibitem [{\citenamefont {Peano}\ and\ \citenamefont
  {Dykman}(2014)}]{Peano_2014}%
  \BibitemOpen
  \bibfield  {author} {\bibinfo {author} {\bibfnamefont {V.}~\bibnamefont
  {Peano}}\ and\ \bibinfo {author} {\bibfnamefont {M.~I.}\ \bibnamefont
  {Dykman}},\ }\href {\doibase 10.1088/1367-2630/16/1/015011} {\bibfield
  {journal} {\bibinfo  {journal} {New J. Phys.}\ }\textbf {\bibinfo {volume}
  {16}},\ \bibinfo {pages} {015011} (\bibinfo {year} {2014})}\BibitemShut
  {NoStop}%
\bibitem [{\citenamefont {Lin}\ \emph {et~al.}(2015)\citenamefont {Lin},
  \citenamefont {Nakamura},\ and\ \citenamefont {Dykman}}]{PhysRevE.92.022105}%
  \BibitemOpen
  \bibfield  {author} {\bibinfo {author} {\bibfnamefont {Z.~R.}\ \bibnamefont
  {Lin}}, \bibinfo {author} {\bibfnamefont {Y.}~\bibnamefont {Nakamura}}, \
  and\ \bibinfo {author} {\bibfnamefont {M.~I.}\ \bibnamefont {Dykman}},\
  }\href {\doibase 10.1103/PhysRevE.92.022105} {\bibfield  {journal} {\bibinfo
  {journal} {Phys. Rev. E}\ }\textbf {\bibinfo {volume} {92}},\ \bibinfo
  {pages} {022105} (\bibinfo {year} {2015})}\BibitemShut {NoStop}%
\bibitem [{\citenamefont {Zhang}\ \emph
  {et~al.}(2017{\natexlab{b}})\citenamefont {Zhang}, \citenamefont {Gosner},
  \citenamefont {Girvin}, \citenamefont {Ankerhold},\ and\ \citenamefont
  {Dykman}}]{ZhangGosnerDykman2017}%
  \BibitemOpen
  \bibfield  {author} {\bibinfo {author} {\bibfnamefont {Y.}~\bibnamefont
  {Zhang}}, \bibinfo {author} {\bibfnamefont {J.}~\bibnamefont {Gosner}},
  \bibinfo {author} {\bibfnamefont {S.~M.}\ \bibnamefont {Girvin}}, \bibinfo
  {author} {\bibfnamefont {J.}~\bibnamefont {Ankerhold}}, \ and\ \bibinfo
  {author} {\bibfnamefont {M.~I.}\ \bibnamefont {Dykman}},\ }\href {\doibase
  10.1103/PhysRevA.96.052124} {\bibfield  {journal} {\bibinfo  {journal} {Phys.
  Rev. A}\ }\textbf {\bibinfo {volume} {96}},\ \bibinfo {pages} {052124}
  (\bibinfo {year} {2017}{\natexlab{b}})}\BibitemShut {NoStop}%
\bibitem [{\citenamefont {Y.Zhang}\ and\ \citenamefont
  {M.Dykman}(2019)}]{zhangDykman2019}%
  \BibitemOpen
  \bibfield  {author} {\bibinfo {author} {\bibnamefont {Y.Zhang}}\ and\
  \bibinfo {author} {\bibnamefont {M.Dykman}},\ }\href@noop {} {\bibfield
  {journal} {\bibinfo  {journal} {arXiv:1906.09973}\ } (\bibinfo {year}
  {2019})}\BibitemShut {NoStop}%
\bibitem [{\citenamefont {Denisenko}\ \emph {et~al.}(2016)\citenamefont
  {Denisenko}, \citenamefont {Munyayev},\ and\ \citenamefont
  {Satanin}}]{1742-6596-681-1-012018}%
  \BibitemOpen
  \bibfield  {author} {\bibinfo {author} {\bibfnamefont {M.}~\bibnamefont
  {Denisenko}}, \bibinfo {author} {\bibfnamefont {V.}~\bibnamefont {Munyayev}},
  \ and\ \bibinfo {author} {\bibfnamefont {A.}~\bibnamefont {Satanin}},\ }\href
  {http://stacks.iop.org/1742-6596/681/i=1/a=012018} {\bibfield  {journal}
  {\bibinfo  {journal} {J. Phys. Conf. Ser.}\ }\textbf {\bibinfo {volume}
  {681}},\ \bibinfo {pages} {012018} (\bibinfo {year} {2016})}\BibitemShut
  {NoStop}%
\bibitem [{\citenamefont {Lörch}\ \emph {et~al.}(2019)\citenamefont {Lörch},
  \citenamefont {Zhang}, \citenamefont {Bruder},\ and\ \citenamefont
  {Dykman}}]{loerchDykman2019}%
  \BibitemOpen
  \bibfield  {author} {\bibinfo {author} {\bibfnamefont {N.}~\bibnamefont
  {Lörch}}, \bibinfo {author} {\bibfnamefont {Y.}~\bibnamefont {Zhang}},
  \bibinfo {author} {\bibfnamefont {C.}~\bibnamefont {Bruder}}, \ and\ \bibinfo
  {author} {\bibfnamefont {M.~I.}\ \bibnamefont {Dykman}},\ }\href@noop {}
  {\bibfield  {journal} {\bibinfo  {journal} {arXiv:1904.09628}\ } (\bibinfo
  {year} {2019})}\BibitemShut {NoStop}%
\bibitem [{\citenamefont {Guo}\ \emph {et~al.}(2013)\citenamefont {Guo},
  \citenamefont {Marthaler},\ and\ \citenamefont
  {Sch\"on}}]{PhysRevLett.111.205303}%
  \BibitemOpen
  \bibfield  {author} {\bibinfo {author} {\bibfnamefont {L.}~\bibnamefont
  {Guo}}, \bibinfo {author} {\bibfnamefont {M.}~\bibnamefont {Marthaler}}, \
  and\ \bibinfo {author} {\bibfnamefont {G.}~\bibnamefont {Sch\"on}},\ }\href
  {\doibase 10.1103/PhysRevLett.111.205303} {\bibfield  {journal} {\bibinfo
  {journal} {Phys. Rev. Lett.}\ }\textbf {\bibinfo {volume} {111}},\ \bibinfo
  {pages} {205303} (\bibinfo {year} {2013})}\BibitemShut {NoStop}%
\bibitem [{\citenamefont {Guo}\ and\ \citenamefont
  {Marthaler}(2016)}]{1367-2630-18-2-023006}%
  \BibitemOpen
  \bibfield  {author} {\bibinfo {author} {\bibfnamefont {L.}~\bibnamefont
  {Guo}}\ and\ \bibinfo {author} {\bibfnamefont {M.}~\bibnamefont
  {Marthaler}},\ }\href {http://stacks.iop.org/1367-2630/18/i=2/a=023006}
  {\bibfield  {journal} {\bibinfo  {journal} {New J. Phys.}\ }\textbf {\bibinfo
  {volume} {18}},\ \bibinfo {pages} {023006} (\bibinfo {year}
  {2016})}\BibitemShut {NoStop}%
\bibitem [{\citenamefont {Svensson}\ \emph {et~al.}(2017)\citenamefont
  {Svensson}, \citenamefont {Bengtsson}, \citenamefont {Krantz}, \citenamefont
  {Bylander}, \citenamefont {Shumeiko},\ and\ \citenamefont
  {Delsing}}]{SvenssonShumeikoDelsing2017}%
  \BibitemOpen
  \bibfield  {author} {\bibinfo {author} {\bibfnamefont {I.-M.}\ \bibnamefont
  {Svensson}}, \bibinfo {author} {\bibfnamefont {A.}~\bibnamefont {Bengtsson}},
  \bibinfo {author} {\bibfnamefont {P.}~\bibnamefont {Krantz}}, \bibinfo
  {author} {\bibfnamefont {J.}~\bibnamefont {Bylander}}, \bibinfo {author}
  {\bibfnamefont {V.}~\bibnamefont {Shumeiko}}, \ and\ \bibinfo {author}
  {\bibfnamefont {P.}~\bibnamefont {Delsing}},\ }\href {\doibase
  10.1103/PhysRevB.96.174503} {\bibfield  {journal} {\bibinfo  {journal} {Phys.
  Rev. B}\ }\textbf {\bibinfo {volume} {96}},\ \bibinfo {pages} {174503}
  (\bibinfo {year} {2017})}\BibitemShut {NoStop}%
\bibitem [{\citenamefont {Svensson}\ \emph {et~al.}(2018)\citenamefont
  {Svensson}, \citenamefont {Bengtsson}, \citenamefont {Bylander},
  \citenamefont {Shumeiko},\ and\ \citenamefont {Delsing}}]{180209259}%
  \BibitemOpen
  \bibfield  {author} {\bibinfo {author} {\bibfnamefont {I.-M.}\ \bibnamefont
  {Svensson}}, \bibinfo {author} {\bibfnamefont {A.}~\bibnamefont {Bengtsson}},
  \bibinfo {author} {\bibfnamefont {J.}~\bibnamefont {Bylander}}, \bibinfo
  {author} {\bibfnamefont {V.}~\bibnamefont {Shumeiko}}, \ and\ \bibinfo
  {author} {\bibfnamefont {P.}~\bibnamefont {Delsing}},\ }\href@noop {}
  {\bibfield  {journal} {\bibinfo  {journal} {Appl. Phys. Lett.}\ }\textbf
  {\bibinfo {volume} {113}},\ \bibinfo {pages} {022602} (\bibinfo {year}
  {2018})}\BibitemShut {NoStop}%
\bibitem [{\citenamefont {Chang}\ \emph {et~al.}(2019)\citenamefont {Chang},
  \citenamefont {Sabin}, \citenamefont {Forn-Diaz}, \citenamefont {Quijandria},
  \citenamefont {Vadiraj}, \citenamefont {Nsanzineza}, \citenamefont
  {Johansson},\ and\ \citenamefont {Wilson}}]{SandboWilson2019}%
  \BibitemOpen
  \bibfield  {author} {\bibinfo {author} {\bibfnamefont {C.~S.}\ \bibnamefont
  {Chang}}, \bibinfo {author} {\bibfnamefont {C.}~\bibnamefont {Sabin}},
  \bibinfo {author} {\bibfnamefont {P.}~\bibnamefont {Forn-Diaz}}, \bibinfo
  {author} {\bibfnamefont {F.}~\bibnamefont {Quijandria}}, \bibinfo {author}
  {\bibfnamefont {A.}~\bibnamefont {Vadiraj}}, \bibinfo {author} {\bibfnamefont
  {I.}~\bibnamefont {Nsanzineza}}, \bibinfo {author} {\bibfnamefont
  {G.}~\bibnamefont {Johansson}}, \ and\ \bibinfo {author} {\bibfnamefont
  {C.}~\bibnamefont {Wilson}},\ }\href@noop {} {\bibfield  {journal} {\bibinfo
  {journal} {arXiv:1907.08692}\ } (\bibinfo {year} {2019})}\BibitemShut
  {NoStop}%
\bibitem [{\citenamefont {Heugel}\ \emph {et~al.}(2019)\citenamefont {Heugel},
  \citenamefont {Oscity}, \citenamefont {Eichler}, \citenamefont {Zilberberg},\
  and\ \citenamefont {Chitra}}]{heugel2019}%
  \BibitemOpen
  \bibfield  {author} {\bibinfo {author} {\bibfnamefont {T.~L.}\ \bibnamefont
  {Heugel}}, \bibinfo {author} {\bibfnamefont {M.}~\bibnamefont {Oscity}},
  \bibinfo {author} {\bibfnamefont {A.}~\bibnamefont {Eichler}}, \bibinfo
  {author} {\bibfnamefont {O.}~\bibnamefont {Zilberberg}}, \ and\ \bibinfo
  {author} {\bibfnamefont {R.}~\bibnamefont {Chitra}},\ }\href@noop {}
  {\bibfield  {journal} {\bibinfo  {journal} {arXiv:1903.02311}\ } (\bibinfo
  {year} {2019})}\BibitemShut {NoStop}%
\bibitem [{\citenamefont {Dykman}(2007)}]{PhysRevE.75.011101}%
  \BibitemOpen
  \bibfield  {author} {\bibinfo {author} {\bibfnamefont {M.~I.}\ \bibnamefont
  {Dykman}},\ }\href {\doibase 10.1103/PhysRevE.75.011101} {\bibfield
  {journal} {\bibinfo  {journal} {Phys. Rev. E}\ }\textbf {\bibinfo {volume}
  {75}},\ \bibinfo {pages} {011101} (\bibinfo {year} {2007})}\BibitemShut
  {NoStop}%
\bibitem [{\citenamefont {Sondhi}\ \emph {et~al.}(1997)\citenamefont {Sondhi},
  \citenamefont {Girvin}, \citenamefont {Carini},\ and\ \citenamefont
  {Shahar}}]{RevModPhys.69.315}%
  \BibitemOpen
  \bibfield  {author} {\bibinfo {author} {\bibfnamefont {S.~L.}\ \bibnamefont
  {Sondhi}}, \bibinfo {author} {\bibfnamefont {S.~M.}\ \bibnamefont {Girvin}},
  \bibinfo {author} {\bibfnamefont {J.~P.}\ \bibnamefont {Carini}}, \ and\
  \bibinfo {author} {\bibfnamefont {D.}~\bibnamefont {Shahar}},\ }\href
  {\doibase 10.1103/RevModPhys.69.315} {\bibfield  {journal} {\bibinfo
  {journal} {Rev. Mod. Phys.}\ }\textbf {\bibinfo {volume} {69}},\ \bibinfo
  {pages} {315} (\bibinfo {year} {1997})}\BibitemShut {NoStop}%
\bibitem [{\citenamefont {Rose}\ \emph {et~al.}(2016)\citenamefont {Rose},
  \citenamefont {Macieszczak}, \citenamefont {Lesanovsky},\ and\ \citenamefont
  {Garrahan}}]{PhysRevE.94.052132}%
  \BibitemOpen
  \bibfield  {author} {\bibinfo {author} {\bibfnamefont {D.~C.}\ \bibnamefont
  {Rose}}, \bibinfo {author} {\bibfnamefont {K.}~\bibnamefont {Macieszczak}},
  \bibinfo {author} {\bibfnamefont {I.}~\bibnamefont {Lesanovsky}}, \ and\
  \bibinfo {author} {\bibfnamefont {J.~P.}\ \bibnamefont {Garrahan}},\ }\href
  {\doibase 10.1103/PhysRevE.94.052132} {\bibfield  {journal} {\bibinfo
  {journal} {Phys. Rev. E}\ }\textbf {\bibinfo {volume} {94}},\ \bibinfo
  {pages} {052132} (\bibinfo {year} {2016})}\BibitemShut {NoStop}%
\bibitem [{\citenamefont {Macieszczak}\ \emph {et~al.}(2016)\citenamefont
  {Macieszczak}, \citenamefont {Gu\ifmmode \mbox{\c{t}}\else
  \c{t}\fi{}\ifmmode~\u{a}\else \u{a}\fi{}}, \citenamefont {Lesanovsky},\ and\
  \citenamefont {Garrahan}}]{PhysRevLett.116.240404}%
  \BibitemOpen
  \bibfield  {author} {\bibinfo {author} {\bibfnamefont {K.}~\bibnamefont
  {Macieszczak}}, \bibinfo {author} {\bibfnamefont {M.}~\bibnamefont
  {Gu\ifmmode \mbox{\c{t}}\else \c{t}\fi{}\ifmmode~\u{a}\else \u{a}\fi{}}},
  \bibinfo {author} {\bibfnamefont {I.}~\bibnamefont {Lesanovsky}}, \ and\
  \bibinfo {author} {\bibfnamefont {J.~P.}\ \bibnamefont {Garrahan}},\ }\href
  {\doibase 10.1103/PhysRevLett.116.240404} {\bibfield  {journal} {\bibinfo
  {journal} {Phys. Rev. Lett.}\ }\textbf {\bibinfo {volume} {116}},\ \bibinfo
  {pages} {240404} (\bibinfo {year} {2016})}\BibitemShut {NoStop}%
\end{thebibliography}%
\bibliographystyle{apsrev4-1}

%%%%%%%%%%%%%%%%%%%%%%%%%%%%%%%%%%%%%%%%%%%%%%%%%%%%%%
%%%%%%%%%% written text parts %%%%%%%%%%%%%%%%%%%%%%%%%%%%%%%%%%%
%%%%%%%%%%%%%%%%%%%%%%%%%%%%%%%%%%%%%%%%%%%%%%%%%%%%%%

%%%%%%%%%%%%%%%%%%%%%%%%%%%%%%%%%%%%%%%%%%%%%%%%%%%%%%%%%%%%%%%%%%%%%%%%%%%%%%%%%%%%%

\end{document}